\documentclass[twocolumn,showpacs,preprintnumbers,amsmath,amssymb]{revtex4-1}

\usepackage{graphicx}
\usepackage{dcolumn}
\usepackage{bm}
\usepackage{color}

\def\kznn{$K_L^0\rightarrow\pi^0\nu\bar{\nu}$}
\def\kznns{$K_L^0\rightarrow\pi^0\nu\bar{\nu}\;$}
\def\kpitwo{$K_L^0\rightarrow\pi^0\pi^0$}
\def\kpitwos{$K_L^0\rightarrow\pi^0\pi^0\;$}
\def\kpithree{$K_L^0\rightarrow\pi^0\pi^0\pi^0$}
\def\kpithrees{$K_L^0\rightarrow\pi^0\pi^0\pi^0\;$}

\def\kgg{$K_L^0\rightarrow\gamma\gamma$}
\def\kggs{$K_L^0\rightarrow\gamma\gamma\;$}

\def\kpnns{$K^+\rightarrow\pi^+\nu\bar{\nu}\;$}
\def\chargedkpithree{$K_L^0\rightarrow\pi^+\pi^-\pi^0$}
\def\chargedkpithrees{$K_L^0\rightarrow\pi^+\pi^-\pi^0\;$}

\def\semileptonics{$K_L^0\rightarrow\pi l\nu\;$}

\def\kpies{$K_L^0\rightarrow\pi^- e^+\nu\;$}

\def\piz{$\pi^0$}
\def\pizs{$\pi^0\;$} 
\def\zvtx{$Z_{\textrm{VTX}}$}
\def\zvtxs{$Z_{\textrm{VTX}}\;$}
\def\kl{$K_L^0$}
\def\kls{$K_L^0\;$}

\def\gams{$\gamma\;$}
\def\pt{$P_T$}
\def\pts{$P_T\;$}

\def\pims{$\pi^-\;$}

\begin{document}

\title{Experimental study of the decay \kznn}

\newcommand*{\PUSAN}{$^1$Department of Physics, Pusan National University, Busan 609-735, Republic of Korea}
\newcommand*{\SAGA}{$^2$Department of Physics, Saga University, Saga 840-8502, Japan}
\newcommand*{\DUBNA}{$^3$Laboratory of Nuclear Problems, Joint Institute for Nuclear Research, Dubna, Moscow Region 141980, Russia}
\newcommand*{\TAIWAN}{$^{4}$Department of Physics, National Taiwan University, Taipei 10617, Republic of China}
\newcommand*{\ARIZONA}{$^{5}$Department of Physics, Arizona State University, Tempe, Arizona 85287, USA}
\newcommand*{\SOKENDAI}{$^{6}$Department of Particle and Nuclear Research, The Graduate University for Advanced Science (SOKENDAI), Tsukuba, Ibaraki 305-0801, Japan}
\newcommand*{\KEK}{$^{7}$Institute of Particle and Nuclear Studies, High Energy Accelerator Research Organization (KEK), Tsukuba, Ibaraki 305-0801, Japan}
\newcommand*{\OSAKA}{$^{8}$Department of Physics, Osaka University, Toyonaka, Osaka 560-0043, Japan }
\newcommand*{\YAMAGATA}{$^{9}$Department of Physics, Yamagata University, Yamagata 990-8560, Japan}
\newcommand*{\IHEP}{$^{10}$Institute of High Energy Physics, Protvino, Moscow Region 142281, Russia}
\newcommand*{\CHICAGO}{$^{11}$Enrico Fermi Institute, University of Chicago, Chicago, Illinois 60637, USA }
\newcommand*{\NDA}{$^{12}$Department of Applied Physics, National Defense Academy, Yokosuka, Kanagawa 239-8686, Japan}
\newcommand*{\KYOTO}{$^{13}$Department of Physics, Kyoto University, Kyoto 606-8502, Japan}
\newcommand*{\RCNP}{$^{14}$Research Center of Nuclear Physics, Osaka University, Ibaragi, Osaka 567-0047, Japan}

\author{
J. K. Ahn$^1$, Y. Akune$^2$, V. Baranov$^3$, K. F. Chen$^4$, J. Comfort$^5$, M. Doroshenko$^{6, a}$, Y. Fujioka$^2$, Y. B. Hsiung$^4$, T. Inagaki$^{6, 7}$, S. Ishibashi$^2$, N. Ishihara$^7$, H. Ishii$^8$, E. Iwai$^8$, T. Iwata$^9$, I. Kato$^9$, S. Kobayashi$^2$, S. Komatsu$^8$, T. K. Komatsubara$^7$, A. S. Kurilin$^3$, E. Kuzmin$^3$, A. Lednev$^{10, 11}$, H. S. Lee$^1$, S. Y. Lee$^1$, G. Y. Lim$^7$, J. Ma$^{11}$, T. Matsumura$^{12}$, A. Moisseenko$^3$, H. Morii$^{13}$, T. Morimoto$^7$, Y. Nakajima$^{13}$, T. Nakano$^{14}$, H. Nanjo$^{13}$, N. Nishi$^8$, J. Nix$^{11}$, T. Nomura$^{13, b}$, M. Nomachi$^8$, R. Ogata$^2$, H. Okuno$^7$, K. Omata$^7$, G. N. Perdue$^{11}$, S. Perov$^3$, S. Podolsky$^{3}$, S. Porokhovoy$^3$, K. Sakashita$^{8, b}$, T. Sasaki$^9$, N. Sasao$^{13}$, H. Sato$^9$, T. Sato$^7$, M. Sekimoto$^7$, T. Shimogawa$^2$, T. Shinkawa$^{12}$, Y. Stepanenko$^3$, Y. Sugaya$^8$, A. Sugiyama$^2$, T. Sumida$^{13, c}$, S. Suzuki$^2$, Y. Tajima$^9$, S. Takita$^9$, Z. Tsamalaidze$^3$, T. Tsukamoto$^{2, d}$, Y. C. Tung$^4$, Y. W. Wah$^{11}$, H. Watanabe$^{11, b}$, M. L. Wu$^4$, M. Yamaga$^{7, 8, e}$, T. Yamanaka$^8$, H. Y. Yoshida$^9$, Y. Yoshimura$^7$, and Y. Zheng$^{11}$
\\
{\rm (E391a collaboration)}
}

\affiliation{
\PUSAN \\
\SAGA \\
\DUBNA \\
\TAIWAN \\
\ARIZONA \\
\SOKENDAI \\
\KEK \\
\OSAKA \\
\YAMAGATA \\
\IHEP \\
\CHICAGO \\
\NDA \\
\KYOTO \\
\RCNP \\
}

\date{\today}

\begin{abstract}

The first dedicated search for 
the rare neutral-kaon decay 
\kznns has been carried out in the E391a experiment at the KEK 12-GeV 
proton synchrotron. 
The final upper limit of 2.6 $\times 10^{-8}$ at the 90\% confidence level
was set on the branching ratio for the decay.


\end{abstract}


\maketitle


\section{Introduction}
\par
The study of rare kaon decays \cite{kdecay-rare} has played an important role 
in the establishment of the standard model (SM) in particle physics. 
It has also been crucial for understanding the phenomenon of CP violation \cite{kdecay-CP}. 
The rare decay \kznns \cite{litt89,isidori,buras} is a direct CP violation process caused by
a flavor-changing neutral current (FCNC) with transition from a strange to down quark.
\par
The unique characteristic of the \kznns decay is that the branching ratio can be 
calculated with very small theoretical uncertainties.  In the SM
based on the Cabibbo-Kobayashi-Maskawa (CKM) matrix \cite{ckm} for quark flavor mixing, 
the \kznns branching ratio is predicted to be $(2.49\pm 0.39) \times 10^{-11}$ \cite{mescia}. 
The uncertainty of the prediction is dominated by the
allowed range of the imaginary part of a CKM matrix element $V_{td}$, 
which is determined by other measurements,
and the intrinsic theoretical uncertainty is only 1-2\% \cite{buras}.
\par
The determination of CKM matrix elements has been greatly improved
in the past ten years
by measuring various B-decay properties \cite{bdecay}; 
all measurements were consistent with each other within the standard model parameterization. 
By measuring the \kznns decay precisely, we can check the consistency against the 
currently predicted value in the SM.
A deviation indicates new physics beyond the SM 
because of the very small uncertainty in deriving the imaginary part of $V_{td}$
from the branching ratio of the \kznns decay.
\par
The decay \kznns is one of the processes expected to have a significant impact on 
new physics searches because it is an FCNC process that can proceed through loop diagrams, 
including the interactions at short distance and large mass scales.
Grossman and Nir \cite{grossman} pointed out the importance of studying new physics by using 
both the \kznns and \kpnns  decays, where the decay \kpnns is another rare kaon decay with 
small theoretical uncertainties \cite{buras,mescia,brod}. 
Various new physics models have been developed and used to predict 
the \kznns and \kpnns branching ratios \cite{buras, isidori,mesciaweb}.
\par 
The best upper limit obtained in previous experiments was BR(\kznn) $<$
5.9 $\times$ 10$^{-7}$ (90\% CL) \cite{ktev}. 
The ultimate goal of our experimental study is to determine the \kznns branching ratio 
with an accuracy less than 10\% of the value predicted in the SM.
The goal can be achieved by performing a series of experiments with improved and refined 
detection methods. 
\par
The E391a experiment, which was carried out at the KEK 12-GeV proton synchrotron 
(KEK-PS), is the first step of this approach. The main objectives of E391a were not only 
to investigate the decay with a dedicated detector, but also to test and confirm our basic 
experimental methods for \kznns at the highest possible sensitivity.  
Data collection of the E391a experiment started in February 2004 and continued 
until November 2005, which was one month before the shutdown of the KEK-PS. 
The total running time was about twelve months and it was divided into three periods 
(Run-1, Run-2, and Run-3). 
Early results from the first and second periods were reported in 
Ref.~\cite{run1-oneweek} and Ref.~\cite{run2}, respectively. In this article, we report 
the final results on the \kznns decay as obtained from E391a.

\section{Experimental method}
\subsection{Basic method of \kznns detection}
The key signature of the \kznns decay is detection of exactly two photons and 
nothing else.  
The \pizs decays dominantly into two photons ($\gamma\gamma$), 
and the neutrinos are undetectable.  
Because detection of the incident \kls is difficult, measurements can only be made of 
the energy and position of the two photons in a calorimeter located downstream of the \kls decay region, 
without having direct information of the incident particle.
Kinematic constraints are weak for definitive identification of the decay. 
Instead, the decay must be isolated by eliminating all possible backgrounds. 
\par
Because the signal mode is identified as the final state of two photons and nothing else,
processes that make two or more photons can cause background events.
The \kls decays such as \kpitwos and \chargedkpithrees become backgrounds when
extra photons or charged particles escape detection.
The \kggs decay can be a background source because it has only two photons in the final state,
although it is well suppressed by kinematical constraints.
Another background process is hadronic interactions of beam neutrons 
with the residual gas in the beamline or in detectors near the beam.
Any \pizs or $\eta$ produced in these interactions decay into two photons and
produce backgrounds. 
Hyperons produced at the target can cause backgrounds through processes 
such as $\Lambda \to \pi^0 n$;
these hyperon events are strongly suppressed because most of them decay in a 10 m long neutral beamline.
\par
The most important tool for reducing the background is a hermetic detector
system to detect and veto extra particles. 
Because all of the other \kls decays, except for \kggs, are accompanied with 
at least two additional photons or charged particles, 
the detector should be highly sensitive to photons and charged particles. 
\par
The signature of \kznns can be provided by using a small-diameter \kls beam 
(called a ``pencil beam") and measuring precisely the energy and position of the two photons.
Although \kls flux is reduced with a pencil beam, it has several advantages.  
First, the beam hole at the center of the calorimeter, which compromises hermeticity, can be minimized. 
Secondly, the \kls decay vertex position (\zvtx) can be assumed to be on the beam axis.
The vertex position, which is same as the \pizs decay position due to the short lifetime of \piz,
is obtained from the kinematics of the two photons from its decay (See Sec.\ III.B).
The transverse momentum of the \pizs (\pt) with respect to the \kls 
beam axis is also obtained. 
\par
The signal region for \kznns decay can be defined with \zvtxs and \pt.
Requiring a sufficiently large missing transverse momentum 
eliminates contamination from the \kggs decay and also reduces the contamination 
of other \kls decays. Most of the \kl's decay into multiple particles that have 
low momenta in the \kls rest frame, and hence low \pts in the laboratory frame. 
The \zvtxs should be in the region away from beam counters 
and \pts should be in the range from 120 to 240 MeV/$c$, 
where the maximum momentum of \piz's in the \kl-rest frame is 231 MeV/$c$ for \kznn. 
\par
Production of \piz's through beam-gas interactions in the decay fiducial region are 
reduced by having a high vacuum.  Hit-rates of the beam halo (mostly neutrons) 
with the surrounding detectors are minimized by a sharp collimation of the beam. 
In addition, as discussed in Sec.\ III.A and in the subsequent analysis of the
data, mis-combinations of two photons from background processes and/or 
mis-measurements of energies and positions were found to be major sources 
of backgrounds in the E391a experiment. 
The photons were often produced from interactions of the beam halo and can lead to incorrect determinations of \zvtx.  
Some backgrounds from beam interactions can be reduced by detecting low-energy deposits from recoil 
particles emitted after the interactions.  More detailed descriptions of the 
detection method have been reported in the original proposal of E391a \cite{design}. 
\par
Inefficiencies of photon and charged particle detection provide backgrounds. 
The decay \kpitwos is the most serious background source arising from an inefficiency 
because it has only two extra photons in the final state. 
The inefficiency of photon detection was investigated in a series of experiments 
\cite{phineff}, which showed that the inefficiency monotonically decreased with the 
incident photon energy. 
It was also found that photon detection with a very low energy threshold is necessary 
to achieve a small overall inefficiency, even for high-energy photons. If an extremely 
low energy threshold around 1 MeV was set, the backgrounds from other \kls decays were 
reduced to a negligible level within the sensitivity of the experiment. 
\par
The following sections describe the application of the basic methods in the
E391a experiment.

\subsection{Beamline}
\begin{figure}[htbp]
\begin{center}
\includegraphics[width=.48\textwidth]{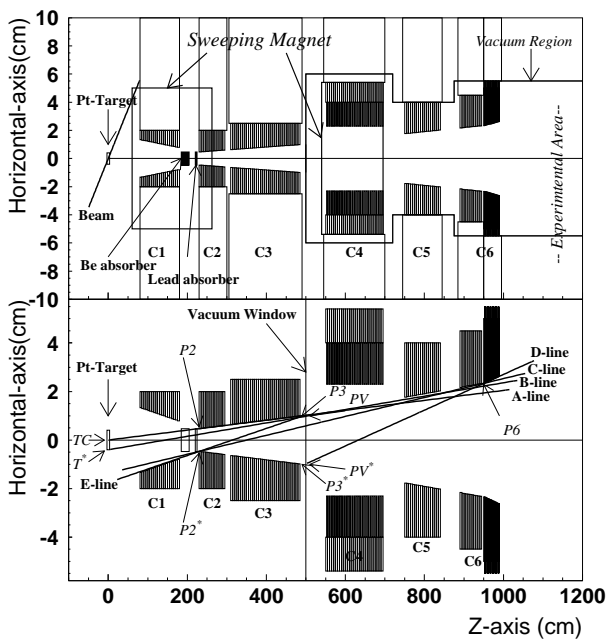}
\end{center}
\caption{Schematic view of the neutral beamline: (top) the arrangements of the components and 
(bottom) the collimation scheme. C1-C3, C5, and C6 were tungsten collimators, assembled as a 
stack of cylindrical blocks 5-cm thick with circular holes of different diameters.
Each collimator approximated the cone-shaped aperture as indicated by the A-E lines 
in the bottom figure.  The $Z = 0$ coordinate is at the center of the production target.}
\label{fig_beamline}
\end{figure}

The production target of the \kls beam was a platinum rod with a length of 60 mm and 
a diameter of 8 mm. The profile of the primary beam was $\sigma$ = 3.3 mm and 
$\sigma$ = 1.1 mm along the $X$ (horizontal) and $Y$ (vertical) axes.
The neutral beam was extracted at an angle of 4$^{\circ}$ with respect to the primary 
proton beam. The target rod and beamline elements were aligned along a straight line 
in the direction of the neutral beam. The total length of the beamline was 10 m. 
The neutral beamline consisted of two dipole magnets (D1, D2) to sweep charged particles 
out of the beam, with field strengths of 4 T$\cdot$m and 3 T$\cdot$m, respectively, and six collimators 
(C1 - C6) to collimate the beam, as shown in Fig.~\ref{fig_beamline}.

\par
Five collimators (C1-C3, C5, and C6) were made of tungsten. 
The C2 and C3 collimators in the upstream end were used to define the beam with the designed 
apertures, which were arranged to form a half cone angle of 2 mrad from the target center.  
The last two collimators, C5 and C6, were used to trim the beam halo.
The most upstream collimator C1 reduced the size of the beam immediately after the target 
without producing a large penumbra.  The total thickness of these five collimators 
was approximately 6 m. A thermal-neutron absorber, which was made of polyethylene terephthalate 
(PET) sheets containing Gd$_2$O$_3$ 40\% in weight, was used for C4. 
The aperture of C4 was set to be larger than  that of the other collimators.
\par
Movable absorbers, made of lead (Pb) and beryllium (Be), were placed between C1 and C2 
to reduce the number of photons and neutrons relative to the \kl's. The absorbers were 
10-mm-diameter rods with the lengths of 5 cm and 30 cm for Pb and Be, respectively. 
The downstream region, starting with a stainless steel window 100 $\mu$m thick
at the upstream end of C4, was evacuated to approximately 1 Pa.
\par
The primary beam on the target was monitored by a secondary-emission chamber (SEC) placed 
upstream of the target, and a target monitor (TM) that was a counter telescope
which viewed the target center at 90 degrees.  
The primary beam position at the target was adjusted with steering magnets 
by monitoring the \kpithrees decay events.
In the $3\pi^0$ events, six photons were detected by a CsI calorimeter
which will be described in Sec.\ II.C.
The center of energy was defined to be $\bm{r}_c = \sum E_i \, \bm{r}_i$, where $E_i$ are the photon 
energies and $\bm{r}_i$ are the photon hit positions at the CsI calorimeter. 
Because the six-photon events were mostly \kpithrees decays, the peak position 
of the center of energy should be on the axis of the beamline.
The position of the neutral beam was maintained to be within 0.2 mm 
from the center throughout the entire running period.
The beam size at the CsI calorimeter, was also monitored by the distribution of the center of energy.
Its diameter was $\sigma = 40$ mm, which was consistent with the beam divergence.
\par
\begin{figure}[htbp]
\begin{center}
\includegraphics[width=.43\textwidth]{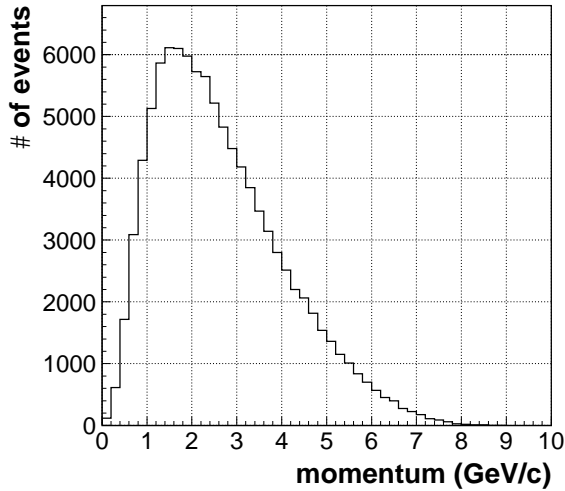}
\end{center}
\caption{
Momentum spectrum of \kls at the exit of C6, obtained from beamline simulations.}
\label{fig_spectrum}
\end{figure}

The peak momentum of the \kls as determined from beamline simulations was 2 GeV/$c$ at the exit of C6, 
as shown in Fig.~\ref{fig_spectrum}.

\begin{figure}[htbp]
\begin{center}
\includegraphics[width=.43\textwidth]{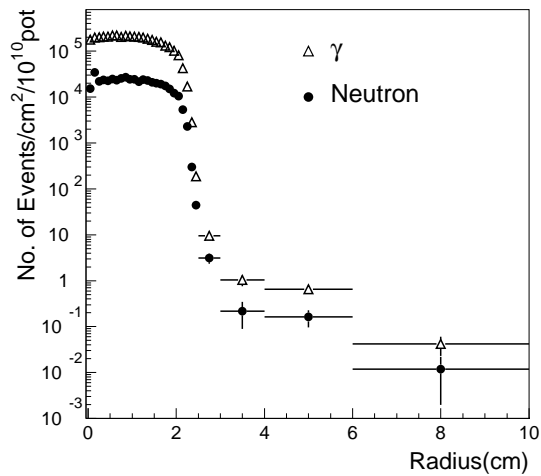}
\end{center}
\caption{
Beam profiles of neutrons and photons above 1 MeV at the exit of C6 collimator, 
obtained from beamline simulations.}
\label{fig_profile}
\end{figure}

\begin{figure}[htbp]
\begin{center}
\includegraphics[width=.43\textwidth]{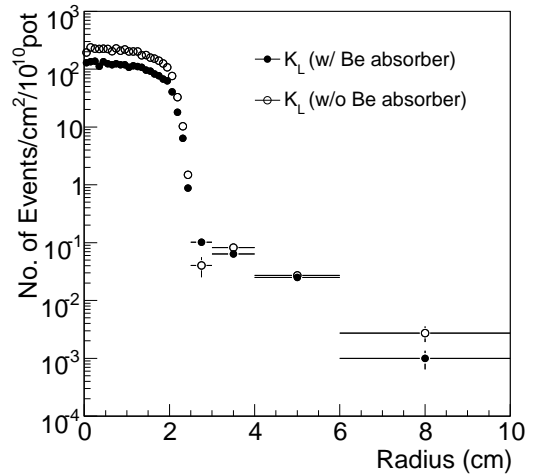}
\end{center}
\caption{
Beam profile of \kls at the exit of C6 collimator, obtained from beamline simulations.
Filled and open circles show the case with and without the Be absorber, respectively.
}
\label{fig_klprofile}
\end{figure}

The initial \pts spread due to the beam divergence of 2 mrad was approximately 4 MeV/$c$. 
The neutron-to-\kls ratio was 60 and the halo-to-core ratio was approximately 10$^{-5}$ 
for both neutrons and photons with energies above 1 MeV, as shown in Fig.~\ref{fig_profile}. 
By inserting a Be absorber, the neutron-to-\kls ratio was reduced to 40, 
with roughly a 45\% loss in \kls flux.
Profiles of the \kls beam are shown in Fig.~\ref{fig_klprofile}.

Punch-through muons were emitted in the direction parallel to the beam axis. 
Their position distribution was almost flat and the flux density was larger than the 
cosmic ray flux by roughly one order of magnitude.
Details of the beamline has been reported elsewhere \cite{beamline}.

\subsection{Detectors}
\begin{figure*}[htbp]
\begin{center}
\rotatebox{-90}{\includegraphics[width=.3\textwidth]{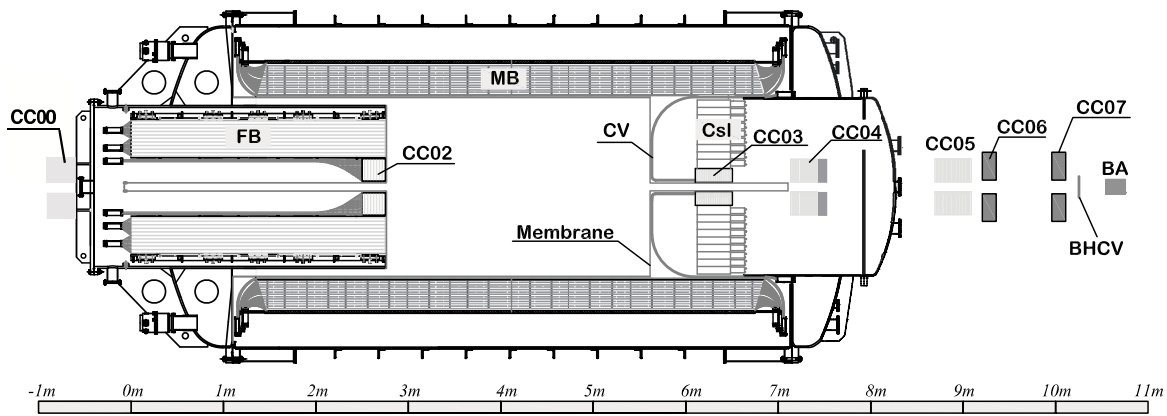}}
\end{center}
\caption{Detection system.}
\label{fig_apparatus}
\end{figure*}

The E391a detection system was located at the end of the beamline.
The detector subsystems were cylindrically arranged around the beam axis, and most of 
them were placed inside a large vacuum vessel, as shown in Fig.~\ref{fig_apparatus}.
From here on, the origin of the coordinate system is defined to be at the upstream end of the 
E391a detector, as shown in Fig.~\ref{fig_apparatus}.
This position was approximately 12 m from the production target.

\subsubsection{CsI calorimeter}
\begin{figure}[htbp]
\begin{center}
\includegraphics[width=.47\textwidth]{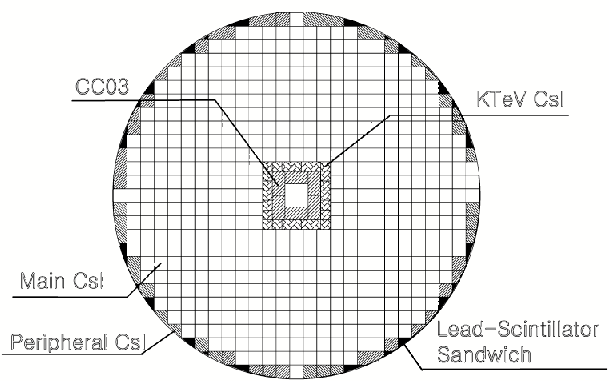}
\end{center}
\caption{Assembly of detector subsystems in the downstream end-cap.}
\label{fig_csi}
\end{figure}

The energy and hit position of photons, such as the two photons
that would come from the \pizs decay in the \kznns signal, were measured by 
using a calorimeter placed at the downstream end of the decay region. 
As shown in Fig.~\ref{fig_csi}, the calorimeter was made of 496 undoped CsI crystals
each with a dimension of 7 $\times$ 7 $\times$ 30 cm$^3$ (Main CsI)~\cite{e162-csi} and 
assembled in a cylindrical shape with an outer diameter of 1.9 m.
There was a 12 cm $\times$ 12 cm beam hole at the center of the calorimeter.
The beam hole was surrounded by a tungsten-scintillator collar counter (CC03) and
24 CsI modules with dimensions of 5 $\times$ 5 $\times$ 50 cm$^3$ (KTeV CsI)~\cite{ktev-csi}.
The gaps at the periphery of the cylinder were filled with 
56 CsI modules with trapezoidal shape and 24 modules 
of lead-scintillator sandwich counters (Lead/Scintillator Sandwich).
These modules were tightly stacked such that the gaps between them were less than 0.1 mm. 
Although the CC03 and the sandwich counters were used only in the veto system, 
the other parts, which consisted of CsI crystals, were used for photon measurements. 
The average light yield of the main CsI modules was 16 pe/MeV, where pe/MeV is the unit 
of the number of photo-electrons emitted from the photomultiplier tube (PMT) cathode 
for an energy deposit of 1 MeV in the detector.  Details of the CsI calorimeter have 
been reported in Ref.~\cite{csi}.

\subsubsection{Charged veto counter}
A set of plastic scintillation counters (named CV) were placed in front of the CsI 
calorimeter to identify events that included the emission of charged particles,
such as $K_L^0 \rightarrow \pi^-e^+\nu\;$ decay~\cite{chineff}.
A total of 32 sector-shaped modules (outer CV) were placed at a distance of 50 cm
from the front face of the CsI calorimeter. 
The outer CV modules were arranged to have overlaps between adjacent modules.
The beam region from the outer CV to the 
CsI was covered by the inner CV, which was a square pipe formed with four plastic scintillator plates. 
The inner and outer CVs were closely connected with aluminum fixtures.
In order to eliminate gaps between the outer and inner modules, 
the inner modules were extended to cover the edge of the outer modules.
The edges of the outer CV and the inner CV were located 
close to the beam, and neutrons in the beam halo frequently interacted with them.

\subsubsection{Barrel counters}
\begin{figure}[htbp]
\begin{center}
\includegraphics[width=.35\textwidth]{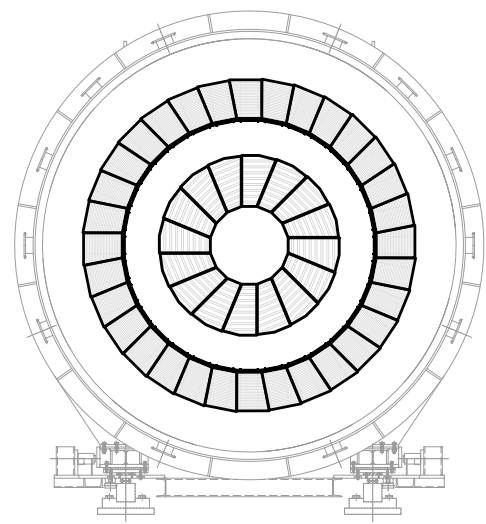}
\end{center}
\caption{Cross-sectional end view of the MB (outer ring) and FB (inner ring).}
\label{fig_barrels}
\end{figure}

The decay region was surrounded with two large lead-scintillator sandwich counters: 
the main barrel (MB) and the front barrel (FB), composed of 32 and 16 modules, 
respectively. The modules were tightly assembled with small gaps \cite{kumitate}, 
as shown in Fig.~\ref{fig_barrels}. 

\begin{figure}[htbp]
\begin{center}
\includegraphics[width=.43\textwidth]{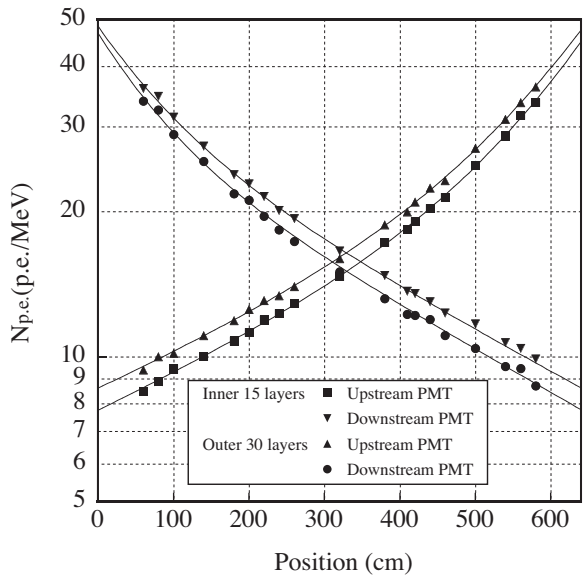}
\end{center}
\caption{Light yield and attenuation for a MB module.}
\label{fig_mblight}
\end{figure}

The lamination was parallel to the beam axis, and the lengths of the MB and FB modules 
were 5.5 m and 2.75 m, respectively. We developed a new type of plastic scintillator
made of a resin mechanically strengthened with a mixture of styrene and methacrylate \cite{ci-scinti}. 
The scintillator counters were fabricated by extrusion.
The thickness of the MB and FB modules were 13.5$X_0$ and 17.2$X_0$, respectively. 
The emitted light was transmitted through wavelength-shifting (WLS) fibers, 
which were glued to each scintillator plate with a pitch of 10 mm.  
The fibers had a double-cladded structure with Y11 as a dopant. 
The green light emitted from Y11 was read by a newly developed photomultiplier tube 
with a high quantum efficiency for green light (EGP-PMT)~\cite{pmt}. 
For readout, each module was divided into an outer and an inner parts. 
The MB module was viewed from both ends, and the FB module was viewed from the upstream end.  
Figure~\ref{fig_mblight} shows the light yields for 4 readout channels of one MB module, 
obtained from a cosmic-ray test. 
The light yield and attenuation of the FB module were similar to those shown in 
Fig.~\ref{fig_mblight} for the MB module. Details of the barrel counters has been 
reported in Ref.~\cite{barrel}.

A layer of plastic scintillator was placed inside of the MB modules to identify charged particles. 
It consisted of 32 modules and was called the barrel charged veto (BCV).
A module was made of a 10-mm-thick plastic scintillator, 
and the light signal was read from both ends by an EGP-PMT through WLS fibers glued 
to the scintillator with a 5-mm pitch.

\subsubsection{Counters close to the beam}
Multiple collar-shaped counters, CC02 -- CC07, were placed along the beam axis.  
Counter CC02 at the entrance of the decay region was a lead-scintillator 
sandwich of the Shashlik type, with WLS fibers piercing the lead and 
scintillator layers.  Counter CC03 at the end of the decay region was a 
tungsten-scintillator sandwich. Counters CC04 -- CC07 covered the solid-angle in 
the downstream direction. Counters CC04 and CC05 were lead-scintillator
sandwiches with WLS fibers glued on each scintillator plate at a pitch of 10 mm, 
which was the same as for the MB and FB.  
Counters CC06 and CC07 were made of SF5 lead-glass.  
The direction of lamination was perpendicular to the beam for 
CC02, CC04, and CC05, and was parallel to the beam for CC03.  The signals 
from CC02 -- CC05 were read by EGP-PMTs. 
\par
Beginning with Run-2, another collar counter called 
CC00 was installed in front of the FB and outside the vacuum vessel to reduce
the effects of halo neutrons. It consisted of 11 layers of a 5-mm-thick plastic 
scintillator interleaved with 10 layers of 20-mm-thick tungsten. 
However, CC00 did not significantly reduce the neutron halo because it was 
not placed as close to the beam as the other collar counters. 
\par
\begin{figure}[htbp]
\begin{center}
\includegraphics[width=.32\textwidth]{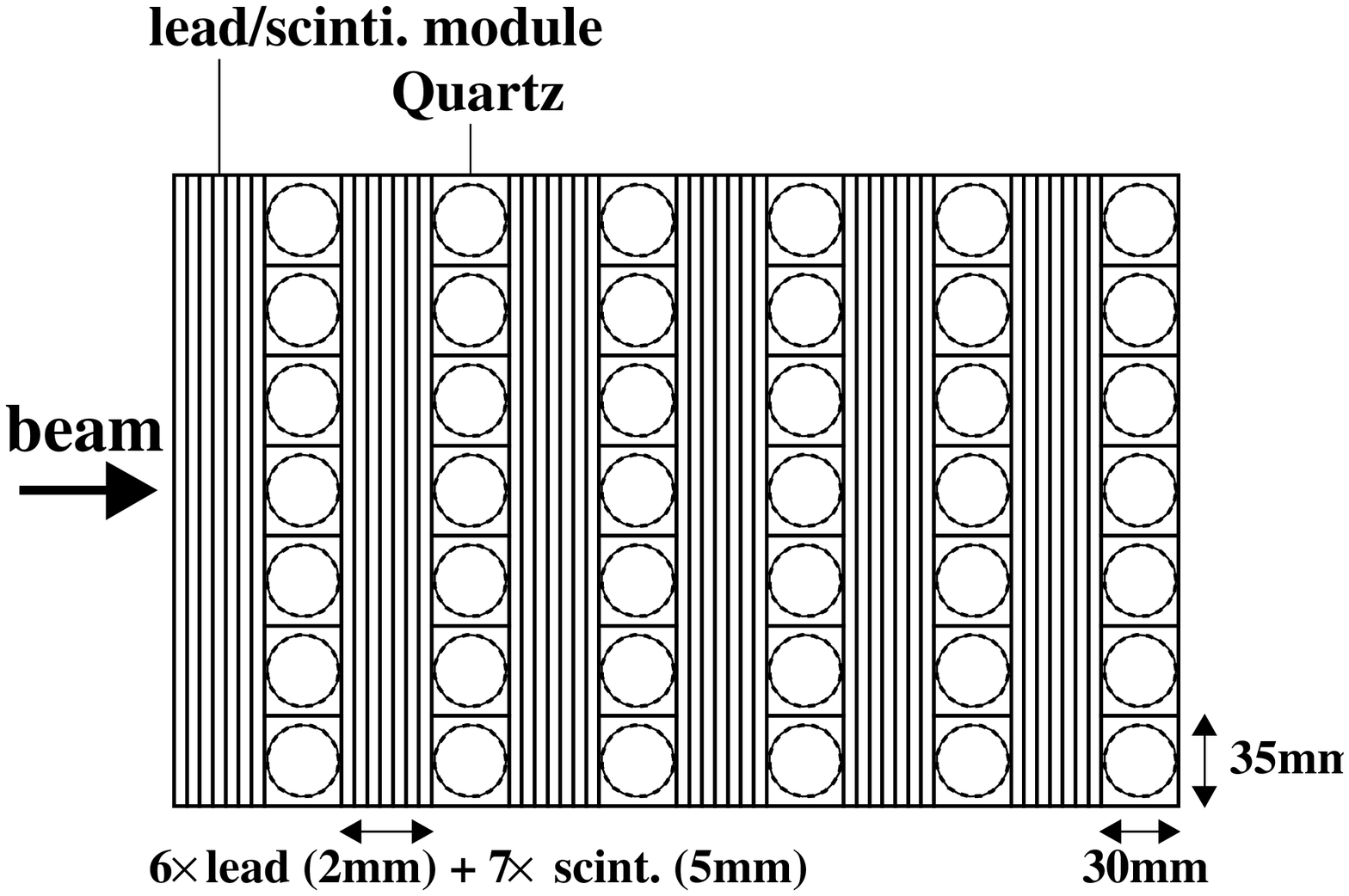} \\
\includegraphics[width=.32\textwidth]{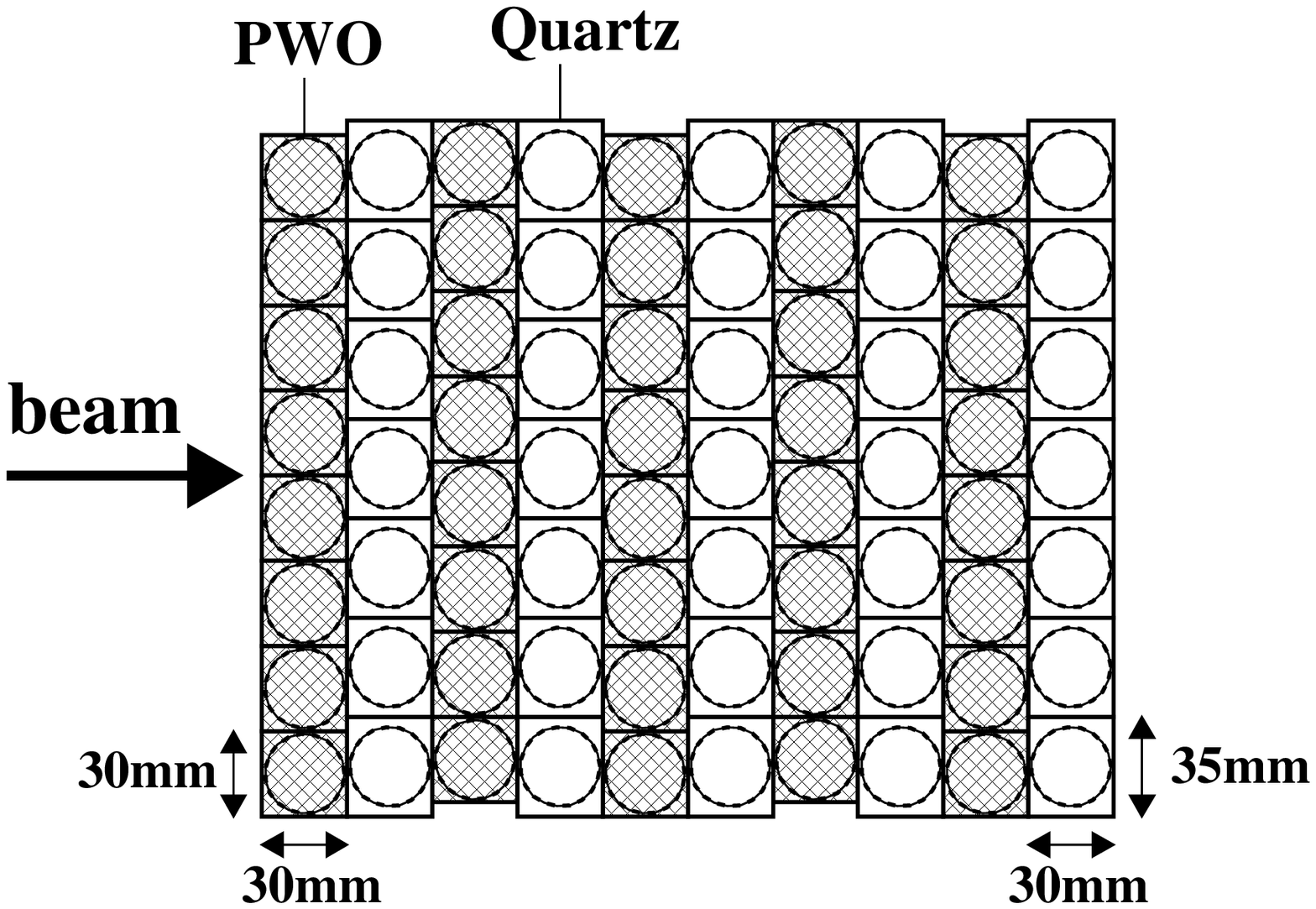}
\end{center}
\caption{BA used in Run-2 (top) and Run-3 (bottom).}
\label{fig_ba}
\end{figure}

A beam-plug counter, back-anti (BA), was placed at the end of detector system along 
the beam axis. In the first two data-taking periods, Run-1 and Run-2, the BA consisted 
of six superlayers, where each superlayer had six plastic-scintillator layers
interleaved with lead sheets, and a single layer of quartz. In Run-3, the lead 
and plastic layers were replaced with PWO crystals as shown in Fig~\ref{fig_ba},
with the intention to separate electromagnetic showers from the neutron hits.
\par
A thin layer of plastic scintillator, the beam hole charged veto (BHCV), 
was placed in front of the BA. It was effective in removing the 
$K_L^0\rightarrow\pi^0\pi^+\pi^-\;$ decay. 
The thickness of the BHCV was 1~mm for Run-1, and 3~mm for Run-2 and Run-3.

\par
The upstream counters, CC02 -- CC04, were placed inside the vacuum vessel, 
and the downstream CC05 -- BA were placed outside the vessel. A vacuum duct, 
which was directly connected to the vacuum vessel, penetrated CC05 -- CC07 
and the vacuum region was extended up to the front face of BHCV.
\par   
The size and basic parameters of CC02 -- CC07, BHCV, and BA are summarized in Table~\ref{tab:table1}.

\begin{table*}
\caption{%
\label{tab:table1}
Size and basic parameters of detectors along the beam.
The origin of the $z$ position corresponds to the start of the E391a detector.
In the sizes of cross section and hole, dia. and sq. represents diameter and square, respectively.
Visible fraction ($R_{vis}$) is defined as the ratio of the energy deposit in 
the active material to that in the whole volume. 
The thickness is expressed in units of radiation length.
The recorded number of photo-electrons per 1-MeV visible energy deposit 
was 10 or more for sandwich counters (CC02-05), and 
at least 0.5 for lead glass detectors (CC06-07), respectively.}
\begin{ruledtabular}
\begin{tabular}{lccclccc}
detector & z position (cm) & outer dimension (cm) & inner dimension (cm) 
& configuration & $R_{vis}$ & thickness ($X_0$) \\
\hline
CC02 & 239.1 & 62.0~dia. & 15.8~dia. 
& lead/scint. & 0.32 & 15.7 \\
CC03 & 609.8 & 25.0~sq. & 12.0~sq. 
& tungsten/scint & 0.23 & 7.6\footnotemark[1] \\
CC04 & 710.3 & 50.0~sq. & 12.6~sq. 
& lead/scint. & 0.28 & 11.8 \\
CC05 & 874.1 & 50.0~sq. & 12.6~sq. 
& lead/scint. & 0.28 & 11.8 \\
CC06 & 925.6 & 30.0~sq. & 15.0~sq. 
& lead glass & 1.0 & 6.3  \\
CC07 & 1000.6 & 30.0~sq. & 15.0~sq. 
& lead glass & 1.0 & 6.3 \\
BHCV & 1029.3 & 23.0~sq. & no
& plastic scint. & 1.0 & 0.007 \\
BA(Run-2) & 1059.3 & 24.5~sq. & no 
      & quartz & 1.0 & 1.5 \\
& & & & lead/scint. & 0.31 & 13.3 \\
BA(Run-3) & 1059.3 & 24.5~sq. & no
      & quartz & 1.0 & 1.2 \\
& & & & PWO & 1.0 & 16.8 \\
\end{tabular}
\end{ruledtabular}
\footnotetext[1]{The lamination of CC03 was parallel to the beam.}
\end{table*}

\subsubsection{Vacuum}
The vacuum pressure in the decay region had to be maintained below 10$^{-5}$ Pa in order 
to reduce the \pizs backgrounds produced by beam-gas interactions to a negligible level 
at the sensitivity corresponding to the SM prediction of \kznn.
The amount of dead material along the path of particles from the decay vertex to the detector had to be 
minimized in order to achieve highly efficient detection even for low-energy particles. 
A differential pumping method was adopted for this purpose. 
\par
\begin{figure}[htbp]
\begin{center}
\includegraphics[width=.3\textwidth]{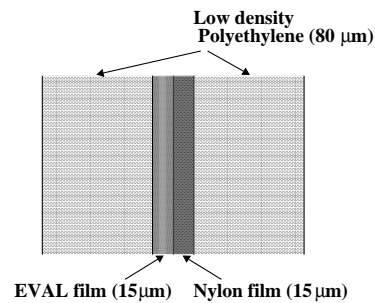}
\end{center}
\caption{
Membrane used for the vacuum separation.
}
\label{fig_membrane}
\end{figure}

The entire detection system was placed in a large vacuum vessel, as shown in
Fig.~\ref{fig_apparatus}.  The pressure in the outer part, where the detectors were placed, 
was around 1 Pa, and the pressure in the inner part, through which the beam passed, was 
$1 \times 10^{-5}$ Pa.  
The two regions were separated by a laminated membrane sheet 
with a thickness of 20 mg/cm$^2$. 
As shown in Fig.~\ref{fig_membrane},
the sheet was a lamination of four films. 
The EVAL film had low transmission for oxygen gas (mostly air) and the nylon film strengthened the sheet. 
The polyethylene layers on both sides were used to make a tight connection by using a heat iron press.
The bag-shaped membrane covered the inner surface of the CsI detectors by using a 
skeleton structure of thin aluminum pipe, similar to a camping tent. 
\par
Two sets of rotary and root pump systems were connected through a manifold
and eight ports to the outer-vacuum part. The pumping speed of each system was 
1200 m$^3$/hour. Four turbo molecular pumps, each having the pumping speed of 800 l/sec, 
were connected between the inner vacuum part and the manifold.  They produced 
the necessary pressure difference of an additional five orders of magnitude.
\par  
For the PMTs installed in vacuum, the pressure had to be less than 10 Pa to prevent 
high voltage discharges.  The PMT operation in vacuum additionally caused a cooling 
problem due to the absence of convection. We modified the configuration of the
resistor chains of the PMTs, and cooled them with a water circulation system 
as described in Ref.~\cite{csi}. The temperature was stable within $\pm 0.1^{\circ}$ 
for the CsI calorimeter.

\subsubsection{Electronics}
About 1000 PMTs were used as readout devices for all detectors. 
Signals from the detectors in the vacuum vessel were extracted through individual 
feed-through connectors. A connector was made of a coaxial cable simply molded to a metal 
flange with a resin. 
\par
All detector signals were fed to Amplifier-Discriminator (AD) modules, 
which were newly developed for the experiment. An AD module accepted 16 PMT signals 
and generated 16 analog signals, 16 discriminated signals, and two 8-channel linear sums.  
Because early discrimination prevents time deterioration due to distortion of the signal 
shape in the cable, the AD modules were placed near the vacuum vessel to shorten the cables.
\par
The analog output signal was derived from each input signal with a throughput of 95$\%$. 
It was sent to a charge-sensitive analog-to-digital converter (ADC) in the counting hut 
through a 90-m coaxial cable, while the cable length of the other outputs was 30 m. 
The discriminator output was generated with a very low threshold of 1 mV, which corresponded 
to an energy deposit of 1 MeV for the CsI calorimeter and below 1 MeV for the other detectors.
The discriminator output was sent to a time-to-digital converter (TDC) in the counting hut 
through 30 m of twisted-pair cable after passing through a fixed delay of 300 ns in the AD 
module. The signal summed over 8 inputs was sent to the counting hut through 30 m of 
coaxial cable and used to form trigger signals for data acquisition. 
\par
The ADC module, LeCroy FASTBUS 1885F, had 96 input channels, 
each channel having the equivalent dynamic range of a 15-bit ADC in its 12-bit data
by using a bi-linear technique.
The typical resolution at a low energy range was
0.13 MeV/bit for the CsI calorimeter, and 0.035 MeV/bit for photon veto detectors.
The gate width for the CsI calorimeter was 200 ns. 
The TDC module in TKO (Tristan KEK Online-system), which was an electronics platform developed in KEK~\cite{tko},
was operated at a 
full range of 200 ns and a resolution of 50 ps. The analog and discriminated signals were 
already delayed by 300 ns (60-m cable with a propagation velocity of 20 cm/ns) 
compared to the timing of linear-sum signal at the entrance of the counting hut. 
All cables to the counting hut were placed inside trays covered with copper-plated iron sheets 
to minimize ground noise caused by alternating magnetic fields. 
The pedestal widths of almost all the ADC channels were less than 1 bit. 
\par
Any failures in the PMTs lead to serious problems in the experiment. Almost all 
PMTs were operated at voltages below 60\% of the rated voltage. Such a large margin was 
made possible by using very sensitive ADCs and very low thresholds for the TDCs. 
All PMTs operated properly during the entire running time.
\par
Multi-hit TDCs (MTDC) were used for the BHCV and BA to cope with high counting rates. 
We did not benefit from using MTDCs in Run-1 because the input pulse width was set 
to 100 ns.  For Run-2 and Run-3, the pulse width was shortened to 25 ns.

\subsection{Trigger and data acquisition}
\begin{figure}[htbp]
\begin{center}
\includegraphics[width=.43\textwidth]{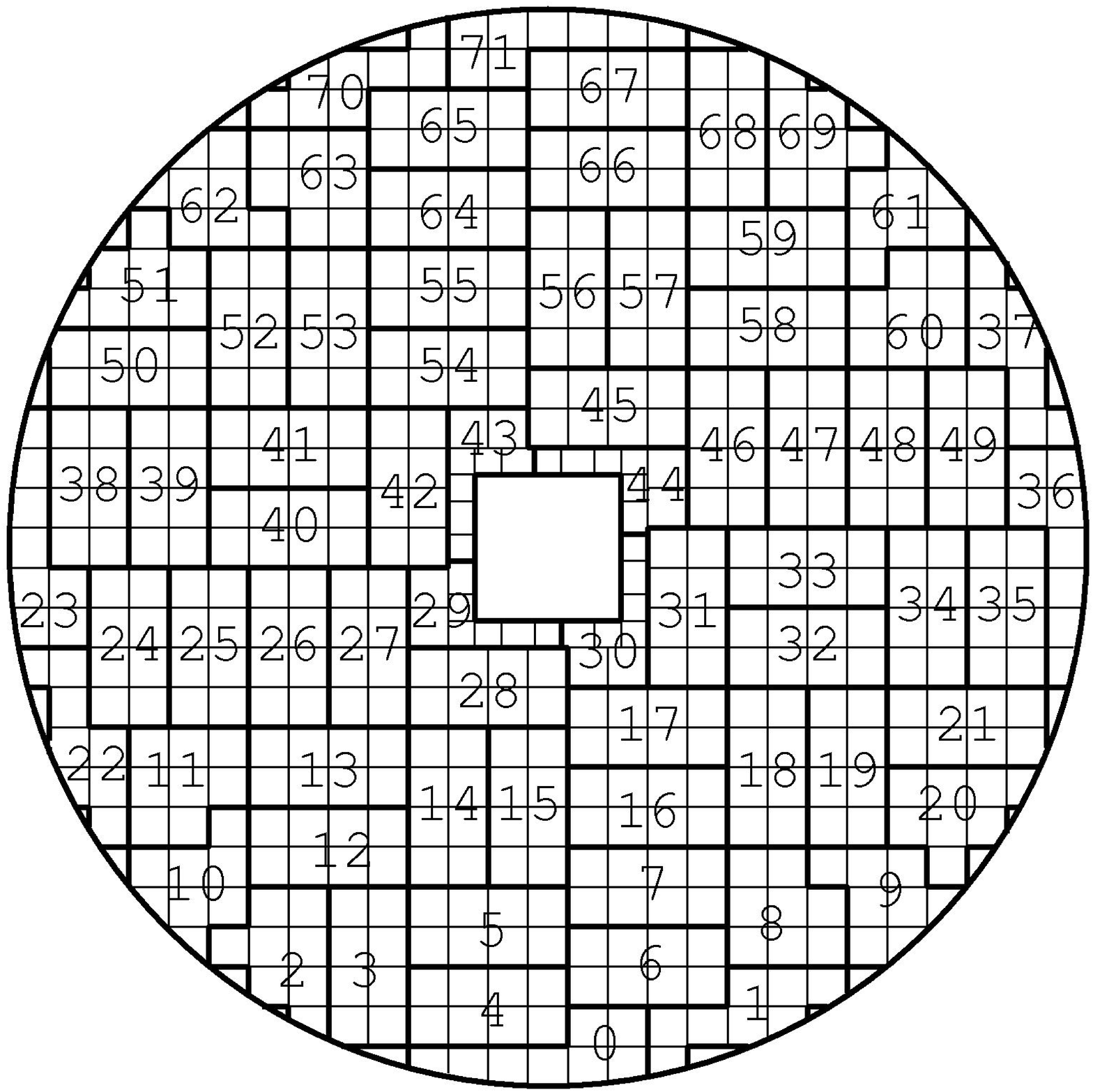}
\end{center}
\caption{Segments of CsI blocks.}
\label{fig_hwclustering}
\end{figure}

The signals from eight adjacent CsI blocks (a segment) were collected into a summed
signal; there were 72 such segments as shown in Fig.~\ref{fig_hwclustering}. 

The segments were used to form a trigger signal.
A threshold, which corresponded to an energy deposit of 80 MeV, was applied to
each summed signal, and the number of segments whose summed signal was above the threshold, 
N$_{\rm HC}$, was counted. 
We required N$_{\rm HC} \ge 2$ for the \kznns trigger.
We also required an anti-coincidence of several veto counters, CV, MB, FB, and CC02 -- CC05 
with a rather high threshold. 
Later, in the offline analysis, these conditions were set tighter: 
a higher threshold for photon detection and lower thresholds for the vetos. 

\par
In addition to the trigger for \kznn, we prepared triggers for light pulsars: 
xenon lamps for CsI and LEDs for the other detectors. They were used to monitor
short-term drifts in the PMT gains. Moreover, they were useful for studying the
effects of beam loading by flashing them within and outside the beam spills. 
Triggers for cosmic ray muons and punch-through muons were also used for 
detector calibration. 
\par
We prepared two types of triggers to record accidental hits in the detector system 
by using an electronic pulse generator and the TM, which was a counter telescope near the target. 
The accidental activities observed in the two types of triggers were consistent with each other
for all detectors except the BA and BHCV. 
While the pulsar trigger had no correlation with the event time and was randomized 
with respect to the asynchronous timing of events, 
the TM trigger reflected intensity variations of the primary beam. 
The TM trigger data were normally used because we observed 
the micro time-structure of the beam extraction.
\par
Accidental losses were estimated by applying the analysis cuts used for 
\kznns to the TM trigger data.  
The estimated value was compared with the value estimated from Monte Carlo simulations 
with imposition of the TM trigger data, and consistency was confirmed. 
\par
Data were collected through multiple parallel systems of the VME crates
operated with a CPU to control them. 
Environmental data such as temperatures at 100 locations, 
vacuum pressure at several places, and single counting rates of 
sampled channels were accumulated by using PCs.
The CPU and PCs were distributed on the network.
The basic software used in the E391a experiment was MIDAS \cite{midas}. 
\par
The dead time of the DAQ system was around 600 $\mu$s/event. 
In a typical run, in which the proton intensity was $2.5\times 10^{12}$ POT (protons on target) 
in a 2-s spill every 4 s, the trigger rate was 300 per spill (150 Hz), 
and the live time was 91\%.  The data size was 3 Mbytes/spill and the typical 
data size collected per day was 60 GB.

\subsection{Calibration}
Good linearity between the energy deposit and the ADC output was observed in all 
detector subsystems. 
Their gains were calibrated through a constant in units of MeV/bit.
The gain constants of all detectors were basically calibrated in situ, 
after assembling and in vacuum, by using cosmic-ray muons and/or punch-through muons 
coming from the upstream region of the primary beamline. While the cosmic-ray muons 
primarily traveled in the downward direction, the punch-through muons were parallel 
to the beam. In the case of sandwich detectors, we selected the muons so that their 
primary direction was perpendicular to the lamination.
\par
\begin{figure}[htbp]
\begin{center}
\includegraphics[width=.43\textwidth]{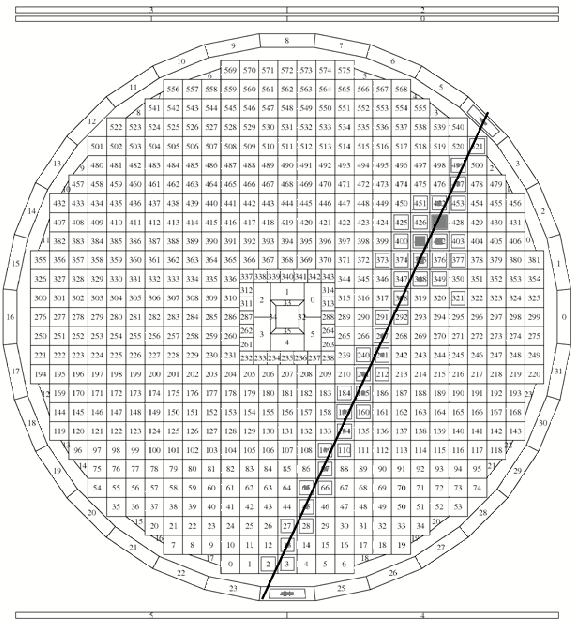}
\end{center}
\caption{Cosmic ray track used for calibration. The outer ring shows the MB.}
\label{fig_cosmic}
\end{figure}

The CsI modules were initially calibrated by using cosmic-ray tracks 
such as the example shown in Fig.~\ref{fig_cosmic}. 

\begin{figure}[htbp]
\begin{center}
\includegraphics[width=.45\textwidth]{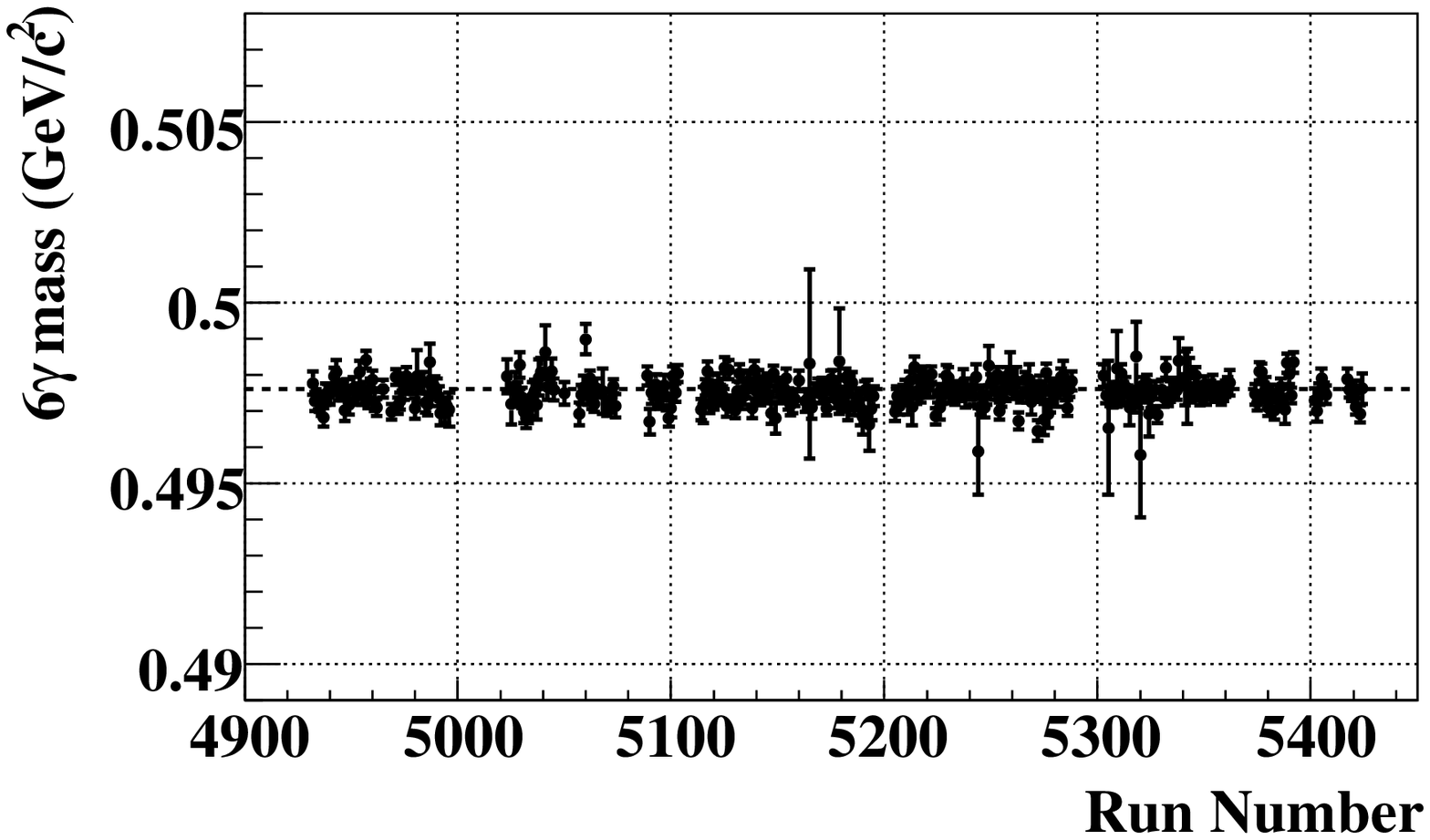} \\
\includegraphics[width=.45\textwidth]{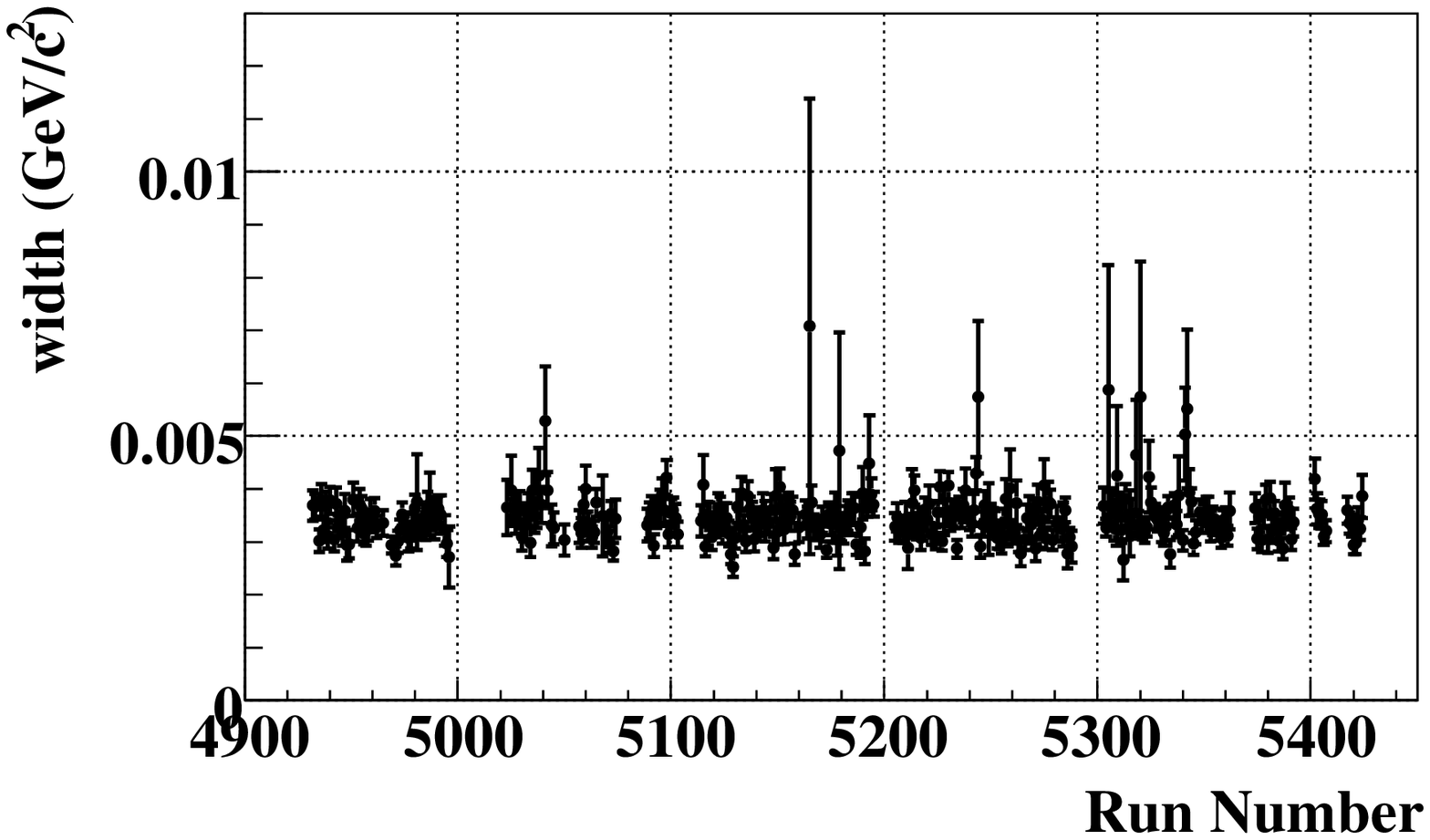}
\end{center}
\caption{
Stability of the kinematic variables over the entire period of Run-3.
The upper and lower graphs show the peak and width of the effective mass distribution 
of 3\pizs from the \kpithrees decay, respectively.}
\label{fig_gainstab}
\end{figure}

The punch-through muons were used to cross-check the cosmic-ray calibration because their 
directions were perpendicular to each other and their penetration lengths were different 
(7 cm for cosmic-ray muons and 30 cm for punch-through muons). The gain constants were 
refined by using the two photons from a \pizs produced in a special run in which an aluminum 
plate was installed on the beam axis. 
Finally, the gain constants were refined by an iteration process based on the
kinematic constraints of \kpithrees decay.
The short-term variation of the gain was corrected by using xenon-lamp light pulses. 
The reconstructed \kls mass and the width were well stabilized, as shown in Fig.~\ref{fig_gainstab}.

\par
In the Monte Carlo simulations, we smeared the energy deposit of each
photon by the function $a/\sqrt{E(\rm{GeV})}+b$.  The first term was given as 
$0.008/\sqrt{E(\rm{GeV})}$, which was consistent with the statistical fluctuation 
of photoelectron yields of 16 pe/MeV. The parameter $b$ was determined to be 
0.004 by tuning it to reproduce the \kpithrees invariant mass distribution.
Because this smearing was applied to the deposited energy instead of the incident 
photon energy,
the small value indicates good understanding of the calibration.
 \par
The beam loading effect was examined by flashing the light pulse during and between 
the beam spills. 
It was negligibly small in the current experiment for all detectors, 
except for the BHCV and BA in which the PMT gain was shifted by 10\%.
\par
The timings were determined relative to one of the photon clusters in the CsI calorimeter. 
First, TDC constants (ns/count) were measured for all TDC channels by using a fixed 
delay, and the time-zero value was calibrated from the data. Cosmic-ray and/or punch-through 
muons were also used for the time-zero calibration. They were determined step-by-step by 
utilizing the overlapping parts of different detectors with respect to the muon tracks.
\par
\begin{figure}[htbp]
\begin{center}
\includegraphics[width=.47\textwidth]{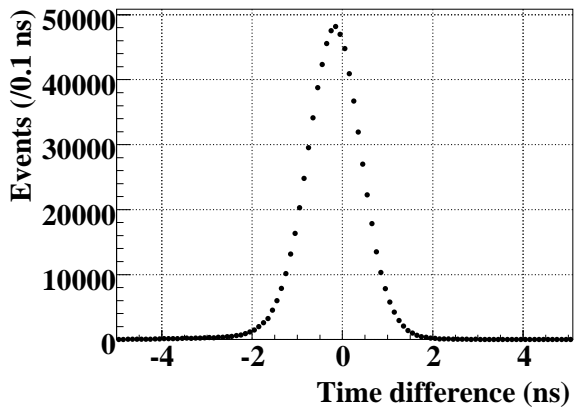}
\end{center}
\caption{Time difference between two photons from the \kpithrees decay,
obtained by subtracting the timing of lower energy photon from that of higher energy photon.
}
\label{fig_csitiming}
\end{figure}

The time-zero values among CsI modules were determined by using cosmic ray tracks.
Finally the calibration was refined by using six 
photons from \kpithrees decays. The time difference between two photons was determined 
with a standard deviation of 0.3 ns, as shown in Fig.~\ref{fig_csitiming}, where the 
timing of a photon cluster was estimated from the timing of the central block that had 
the local maximum energy deposit.

\section{Data analysis}
\subsection{Summary of three runs and outline of present analysis}
During the course of the experiment, there were $(1 - 2.5 ) \times 10^{12}$ 
POT per spill, with a total yield of $4.6 \times 10^{18}$ POT for physics runs.
A summary of three physics runs is listed in Table~\ref{table_runs}.
The experimental setup was modified during intervals between the runs, 
and the running parameters were refined on the basis of the results of a previous run.

\begin{table}[htbp]
\caption{Summary of three physics runs.}
\label{table_runs}
\begin{tabular}{ccccccc}
\hline \hline
Run & \hspace{0.15cm}& Run period & \hspace{0.15cm} & POT & \hspace{0.15cm} & Remarks \\
\hline
Run-1 & & Feb.-Jun. 2004 & & $2.1 \times 10^{18}$ & & Membrane problem \\
Run-2 & & Feb.-Apr. 2005 & & $1.4 \times 10^{18}$ & & Be absorber \\
Run-3 & & Oct.-Dec. 2005 & & $1.1 \times 10^{18}$ & & New BA \\
\hline \hline
\end{tabular}
\end{table}

\par
The data quality of Run-1 was severely affected because the 
membrane for vacuum separation drooped into the beam core, causing many beam interactions. 
The results obtained from a 10\% data sample of Run-1 have been published 
elsewhere \cite{run1-oneweek}. The single-event sensitivity (S.E.S.) was 
$(9.11\pm 0.20_{\textrm{stat}}\pm 0.64_{\textrm{syst.}})\times10^{-8}$ 
with $1.9\pm 1.0$ background events expected in the signal box ($N_{\textrm{bg}}$). 
The complete data sample obtained in Run-1 was analyzed with a  blind analysis. 
In the blind analysis, the signal candidate events were not examined 
until after the final selection criteria were set, and the criteria were not changed later.  
The results of the complete data sample of Run-1 were found to be essentially 
consistent with those obtained from the 10\% data sample. 
The S.E.S. was $(5.14\pm0.25)\times10^{-8}$ with 
N$_{\textrm{bg}}=2.10$ \cite{run1-full}. 
The number of observed events in the signal box (N$_{\rm{ob}}$) 
was 0 for the 10\% sample and 1 for the complete data sample. 
The corresponding upper limits for the branching 
ratio of \kznns mode at the 90\% confidence level (C.L.) were found to be 
$2.1\times10^{-7}$ and $1.26\times10^{-7}$, respectively. 
Because the sensitivity of the Run-1 data sample was not as high as those of 
Run-2 and Run-3, and the running condition of Run-1 was considerably 
different from the later runs, the results of Run-1 are not included 
in the results in this article.
\par
The membrane problem was rectified before Run-2. 
As described in Sec.\ II.C.4,
the plastic scintillator of the BHCV was replaced with a thicker one, 
and the discriminated pulse width for the BHCV and BA was reduced. 
An additional collar counter, CC00, was installed in front of the FB outside 
the vacuum vessel. In Run-2 and Run-3, a neutron absorber made of beryllium 
was inserted into the beam line.  The results for the complete Run-2 data 
sample, S.E.S. = $(2.91\pm 0.31)\times10^{-8}$ with N$_{\textrm{bg}}=0.42\pm 0.14$ 
and N$_{\rm{ob}}=0$, and the upper limit of BR(\kznn) $= 6.7\times10^{-8}$ 
at the 90\% CL, have been published in a letter \cite{run2}.
\par
From the previous Run-2 analysis, the dominant source of the background was found to be 
the interaction of the beam halo (mostly neutrons) with the detectors near the beam 
(``halo neutron background"). The halo neutron background was from three sources:
\piz's produced by the interaction with CC02, and \piz's and $\eta$'s 
produced by the interaction with the CV.  
In the case of CC02-\pizs background, the computed \zvtxs could be shifted downstream due to shower 
leakage or photo-nuclear interactions in the CsI calorimeter. 
In the case of CV-\pizs background, \zvtxs could be shifted upstream due to the fusion 
of multiple photons or the overlap of other hits in the calorimeter. 
In the case of CV-$\eta$ background, 
\zvtxs could be shifted upstream by applying the \pizs assumption in the reconstruction of the two photons.
\par
In the previous Run-2 analysis, the background level N$_{\textrm{bg}}$ was estimated 
with a different method for each background source. A special run with an Al plate was used 
to estimate the background level for CC02-\piz. 
A bifurcation method with a pair of cut sets was used to estimate the CV-\pizs background.
An MC calculation was used to estimate the CV-$\eta$ background.
The difference in estimating these backgrounds made it difficult 
to carry out optimizations with the objective of obtaining the best signal-to-noise ratio, S/N. 
\par
To improve the analysis of Run2-3 data as described in this paper, 
we developed a method to generate a large number of beam halo interactions 
by using the hadron-interaction code FLUKA~\cite{fluka}, whose reliability was confirmed 
by the data from the beam survey \cite{beamline}. 
The simulation generated beam-halo interactions with approximately 8 times 
the total number of real data obtained in Run-3. 
For the CV-$\eta$ background case, we recycled the event seeds 
to boost the $\eta$ production and obtained approximately 80 times larger statistics 
as compared to the Run-3 data.
Although the simulation was carried out for Run-3, the events were
used for Run-2 with minor modifications. The running conditions were almost 
the same in Run-2 and Run-3, except that the super-layers of the lead scintillator 
sandwich of the BA were replaced with PWO hodoscopes in Run-3.
\par
In the analysis presented in this article, we first carefully checked the MC events 
against the data from the aluminum plate run to confirm the hadronic interaction model. 
Next, we optimized the selection criteria by monitoring the S/N ratio of MC data. 
Here, S is the number of events inside the signal region 
(340 $<$ \zvtxs $<$ 500 cm and 120 $<$ \pts $<$ 240 MeV/$c$) 
for the \kznns MC events, and N is the MC background events caused 
by beam-halo interactions~\cite{jiasen}. 
The optimization based on the MC is also effective to avoid potential bias 
in reanalyzing the Run-2 data.
For the data processing, we first masked the events in the signal region 
for both Run-2 and Run-3. 
Once we fixed all the selection criteria by the optimization described above, 
we compared the data and MC for the events outside the signal region. 
After confirming the consistency, we finally examined the events inside the signal region.


\subsection{Event reconstruction}
\subsubsection{Photon clustering}
\begin{figure}[htbp]
\begin{center}
\includegraphics[width=.44\textwidth]{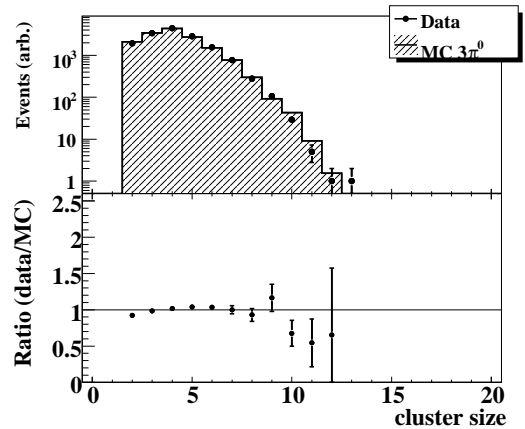}
\end{center}
\caption{
Distribution of the number of crystals having the energy deposit greater than 5 MeV in a photon cluster, 
comparing the data and \kpithrees MC.
The top portion shows a comparison between the number of events in the data and MC, 
and the bottom portion shows the ratio between them.
All analysis cuts for the \kpithrees mode, as will be described in Sec.\ V.B., are imposed.
}
\label{fig_ncluster}
\end{figure}

The energy of each photon was measured by forming a cluster of CsI blocks with finite energy deposits.  
First, a search was made for a block with the locally maximum energy, 
{\em i.e.} the maximum energy among the five blocks geometrically sharing a side. 
An envelope surrounding the local-maximum block was developed 
by gathering blocks that shared one side and had lower energies. 
An envelope with a fixed boundary was required
to reject extra particles that hit the outsides of the two photon clusters. 
It was also necessary to determine that there was only one local maximum within 
the envelope, to discriminate clusters that arose from the fusion of 
more than one photon.
Figure~\ref{fig_ncluster} shows the distribution of the number of crystals contained in one photon cluster
by comparing the six photon events of data and \kpithrees MC.
The number of crystals was obtained by counting the crystals having the energy deposit greater than 5 MeV.
The size of the cluster was well reproduced by the simulation.

\subsubsection{Energy and position correction}
The hit position of a photon at the front of the CsI calorimeter was first estimated 
by determining the center of energy of hit blocks in each photon cluster.
The \zvtxs coordinate was calculated by using two photon clusters and assuming the \pizs mass. 
The energy, position, and injection angle were obtained for each photon. 
\par
\begin{figure}[htbp]
\begin{center}
\includegraphics[width=.44\textwidth]{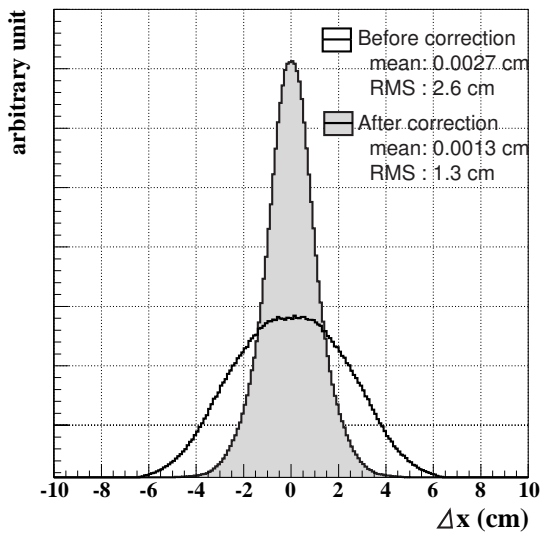}
\end{center}
\caption{
Difference of the reconstructed $x$-position of the photons on the calorimeter
from the true value ($\Delta x$), obtained by \kznns MC simulation. 
The $x$ coordinate denotes the horizontal position at the face of the CsI.
The resolution was improved after the energy and position corrections.
}
\label{fig_hslee_corr}
\end{figure}

In this experiment, the energy and position of each photon 
had to be corrected for the injection angle to obtain better resolution.
The thickness of the CsI crystal (16.2X$_0$) was not sufficient to prevent a 
small leakage of energy from its downstream end, and the energy leakage 
depended on the injection angle. 
We accumulated a large sample of MC data for a single photon with various energies,
positions, and angles, 
and compiled look-up tables to correct energy and position of each photon.
The correction was iteratively carried out by using data from the tables. 
Figure~\ref{fig_hslee_corr} shows a result of the correction.

\subsubsection{Sorting of events}
The events were sorted into several samples according to the number of 
reconstructed photon clusters.
The two-photon sample was used to search for the \kznns decay and for monitoring 
the \kggs decay. The four-photon and six-photon samples were used for monitoring 
the \kpitwos and \kpithrees decays, respectively. 

\subsubsection{\pizs reconstruction}
By using the position and energy of two photon clusters, 
the decay vertex (\zvtx) and the momentum of the \pizs were reconstructed
with assumptions that the invariant mass of two photons was equal to the \pizs mass 
and that the decay vertex was on the beam axis.
The opening angle of two photons ($\theta$) was 
calculated from the equation
\begin{equation}
M_{\pi^0}^2 = 2E_1 E_2 (1 - \cos{\theta}) \:,
\end{equation}
where $E_1$ and $E_2$ are the energies of the two photons. 
The momentum and transverse momentum (\pt) of the \pizs were calculated 
from the decay vertex.

\subsection{Monte Carlo simulations}
\begin{figure}[htbp]
\begin{center}
\includegraphics[width=.47\textwidth]{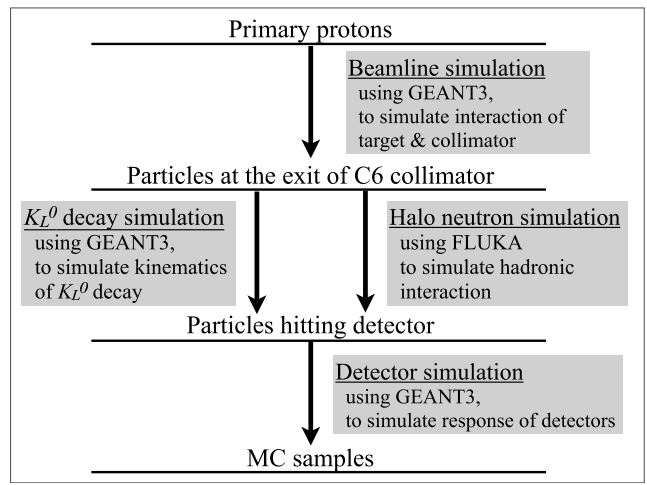}
\end{center}
\caption{
Schematic diagram of simulation.
}
\label{fig_mcscheme}
\end{figure}

We generated Monte Carlo events that had the same structure as the recorded events, 
and analyzed them with the same code as for the real events. 
Because a full shower simulation would require extensive computation time, 
the calculations were separated into several stages and streams, as shown 
in Fig.~\ref{fig_mcscheme}. The first stage was the beamline simulation in 
which the target, collimators, magnetic field, and absorbers were introduced. 
The GEANT3 package~\cite{geant3} with the GFLUKA plug-in code for hadronic 
interactions was used for this simulation. The momentum, position in the 
transverse directions, and angle distributions at the exit of the last 
collimator C6 were obtained for various beam particles.

\par
\begin{figure*}[htbp]
\begin{center}
\includegraphics[width=.44\textwidth]{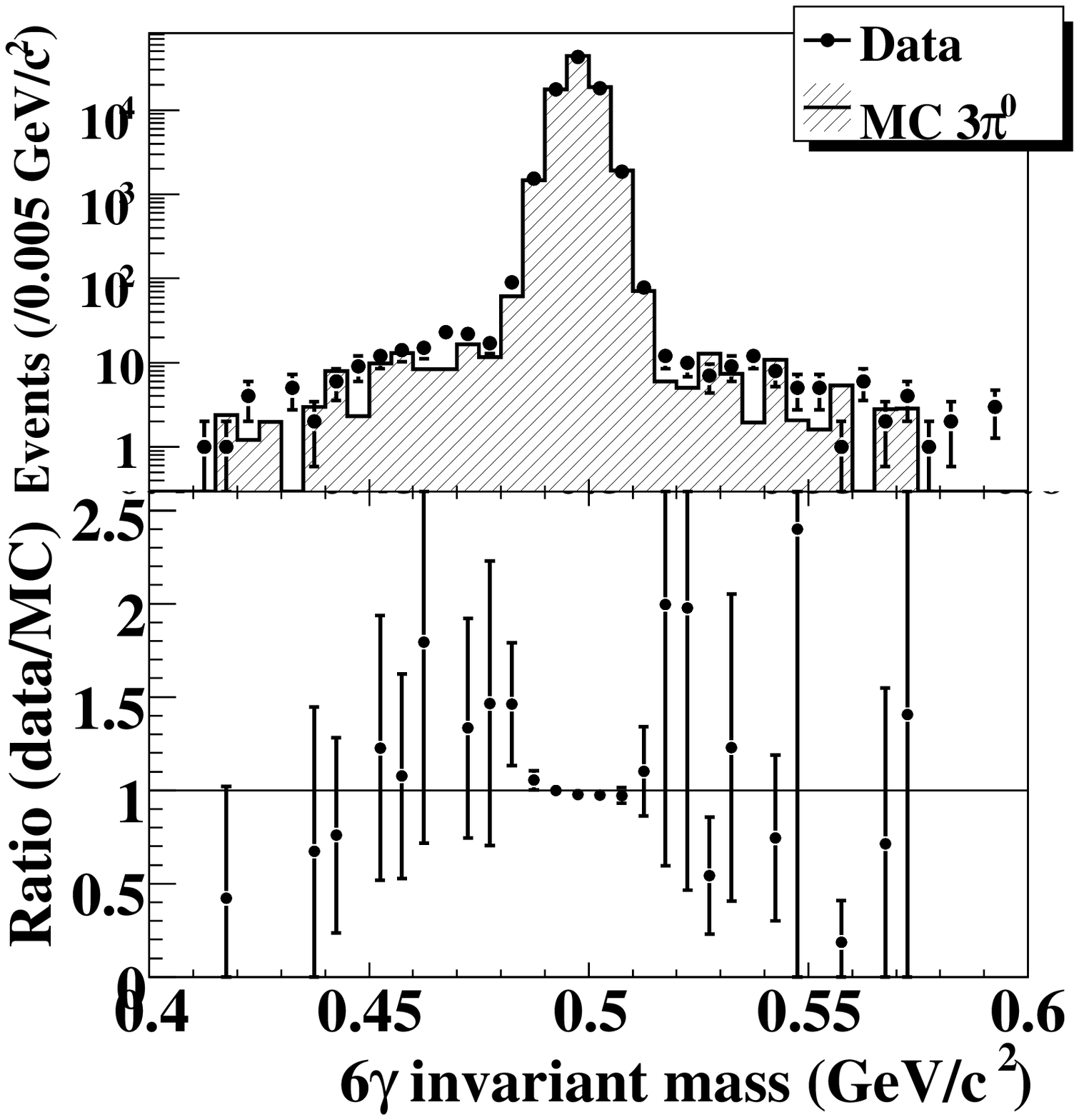}
\includegraphics[width=.44\textwidth]{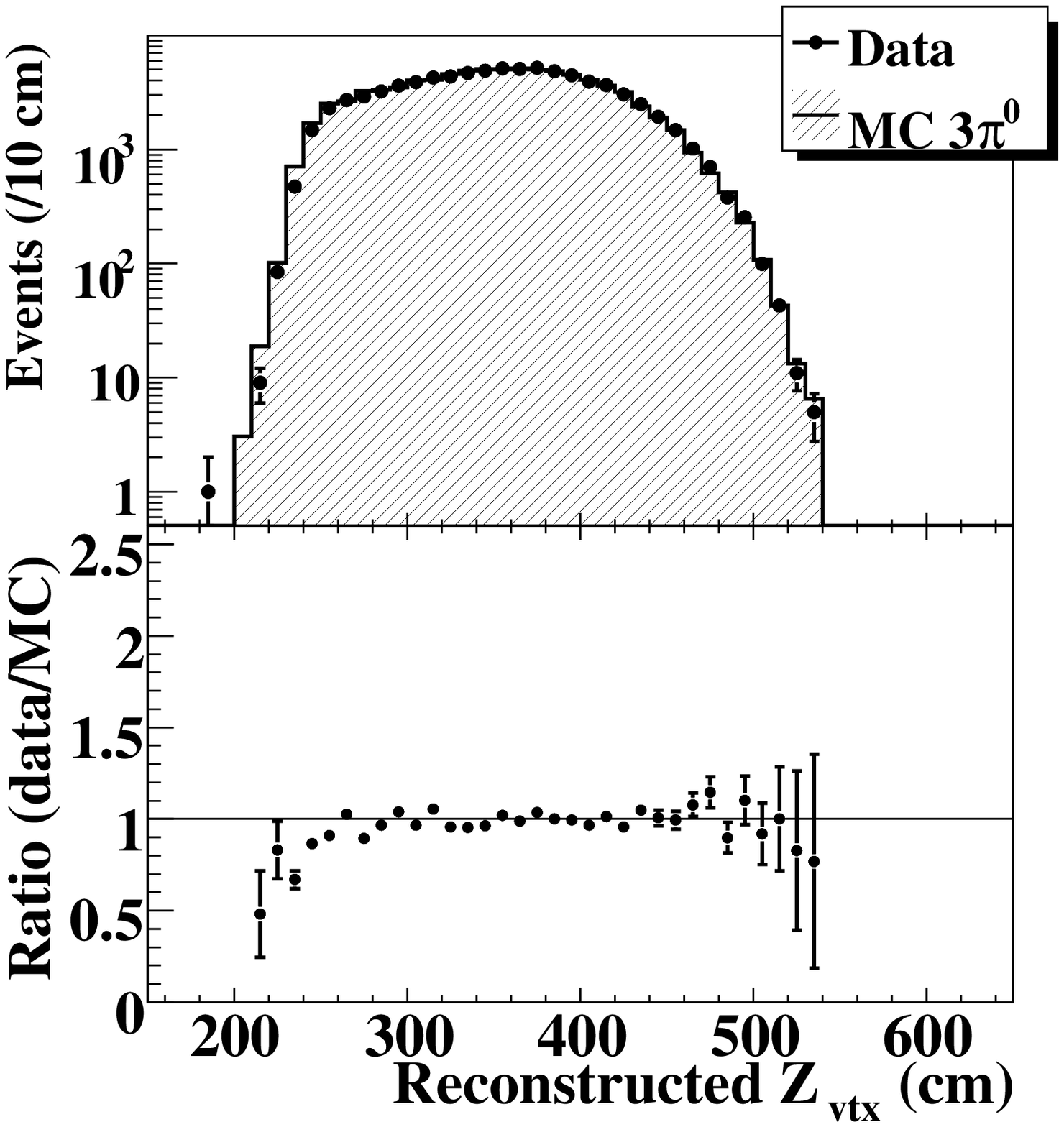} \\
\includegraphics[width=.44\textwidth]{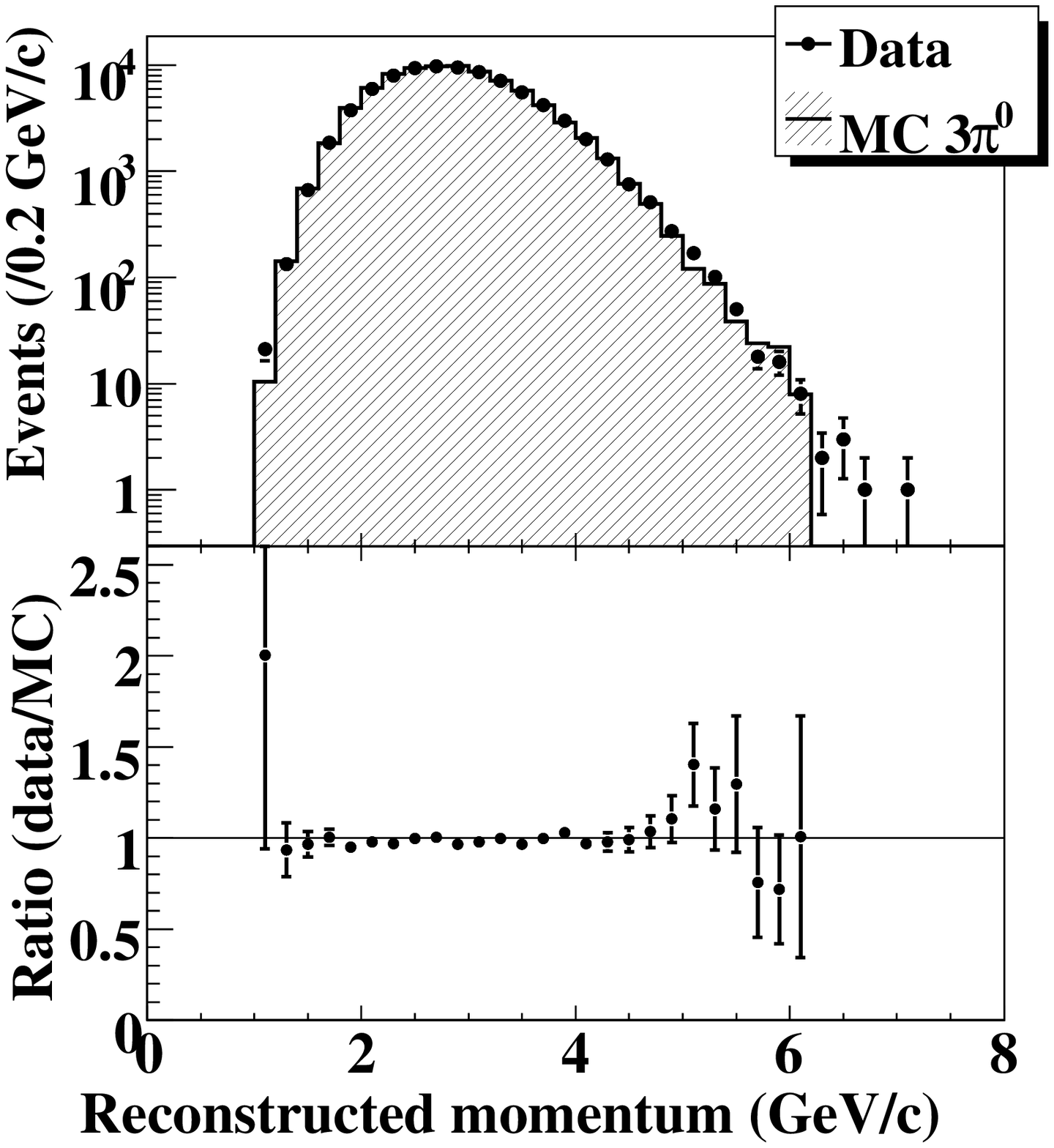}
\includegraphics[width=.44\textwidth]{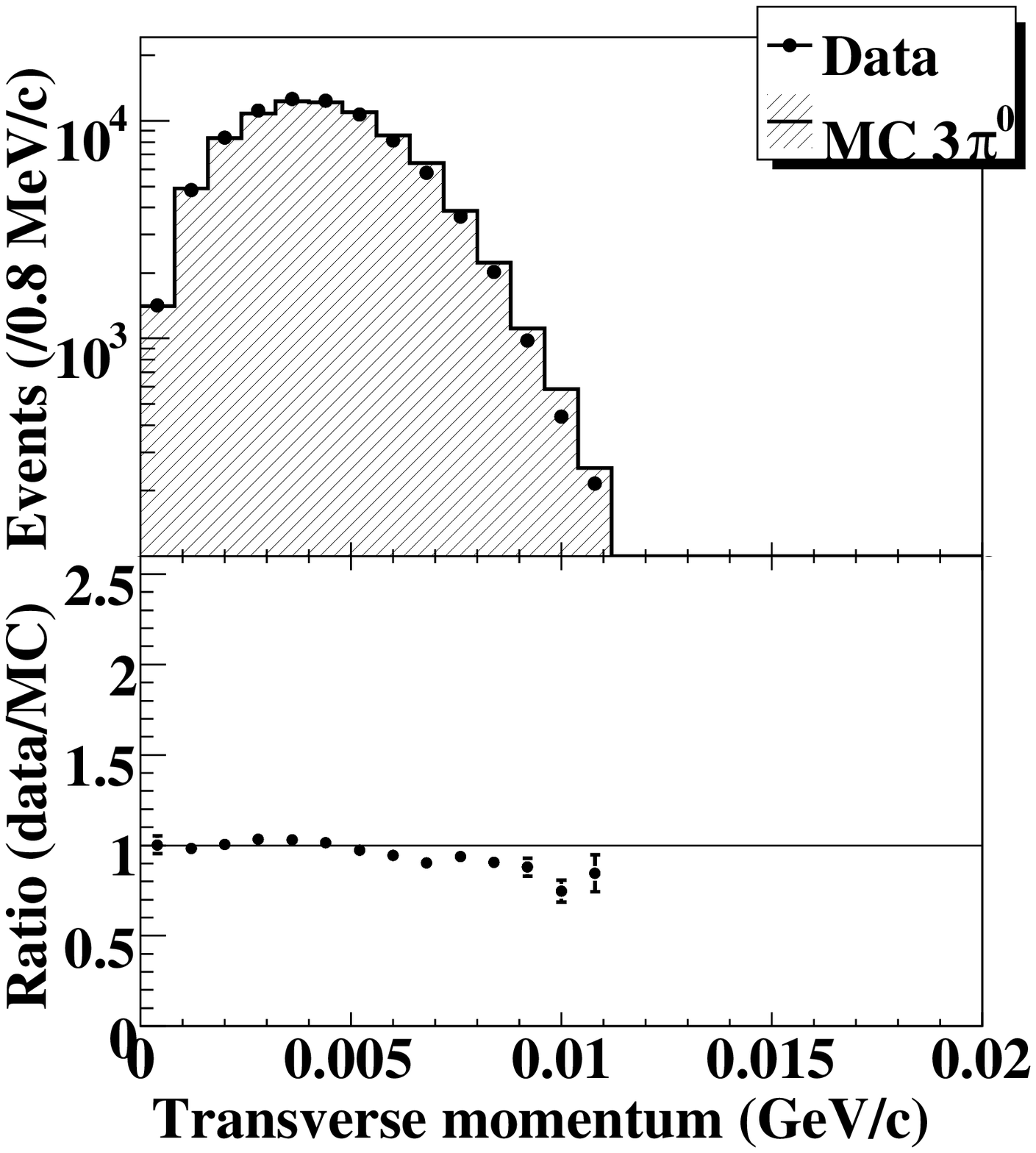}
\end{center}
\caption{
Distributions of reconstructed mass (top left), vertex position (top right), momentum (bottom left), 
and transverse momentum (bottom right) obtained from the 
six-photon data samples, compared to the data and the \kpithrees MC. 
In each plot, the top portion shows a comparison between the number of 
events in the data and MC, and the bottom portion shows the ratio between them.
All the veto cuts and kinematic selections, except those for the 
respective abscissa variable for the \kpithrees mode, are applied in these plots. 
Details of the cuts are described in Sec.\ V.B.}
\label{fig_kpithree}
\end{figure*}

For \kl, functions for the momentum and position 
distributions at the C6 exit were obtained by fitting the Monte Carlo data.
The \kls beam beyond C6 was subsequently 
generated with respect to the momentum spectrum and targeting angle at C6.
The generated \kl's were used for 
the study of various \kls decays through detector simulations with the GEANT3 
package. 
In the generation of the \kznns events, we assumed V-A interactions.
In the detector simulations, particles were traced until their 
energy decreased below the cutoff value (e.g. 0.05 MeV for photons and electrons).
Typical results of \kls decay simulations: reconstructed mass, vertex position,
momentum, and transverse momentum for the six-photon data samples are shown in Fig.~\ref{fig_kpithree}.
Small corrections to the momentum and radial position of the \kls were included.
Each distribution is reproduced by the simulation of the \kpithrees decays.
The method used to reconstruct the six-photon events and 
the event selections are described in a later section.

\par
Another analysis stream from the MC particles at C6 was 
the generation of halo neutrons. 
First, the core neutrons were removed from the neutron data sample. 
The remaining halo neutrons were used multiple times as seeds. 
During the generation, we introduced Gaussian fluctuations
to the momentum, position, and angle of 
each new event to prevent duplications of the same event. 
We used the hadron code FLUKA to simulate the interaction between halo neutrons
and the detector.
Secondary particles generated by FLUKA were fed into the GEANT3 detector simulation.

The detector response was simulated by using the GEANT3 package.
The energy deposit and the timing in each detector subsystem was stored. 
Trigger conditions were simulated according to the energy deposits of the detector elements,
including the segments of the CsI shown in Fig.~\ref{fig_hwclustering}.

\subsection{Reproducibility of the MC simulations}
\begin{figure}[htbp]
\begin{center}
\includegraphics[width=.47\textwidth]{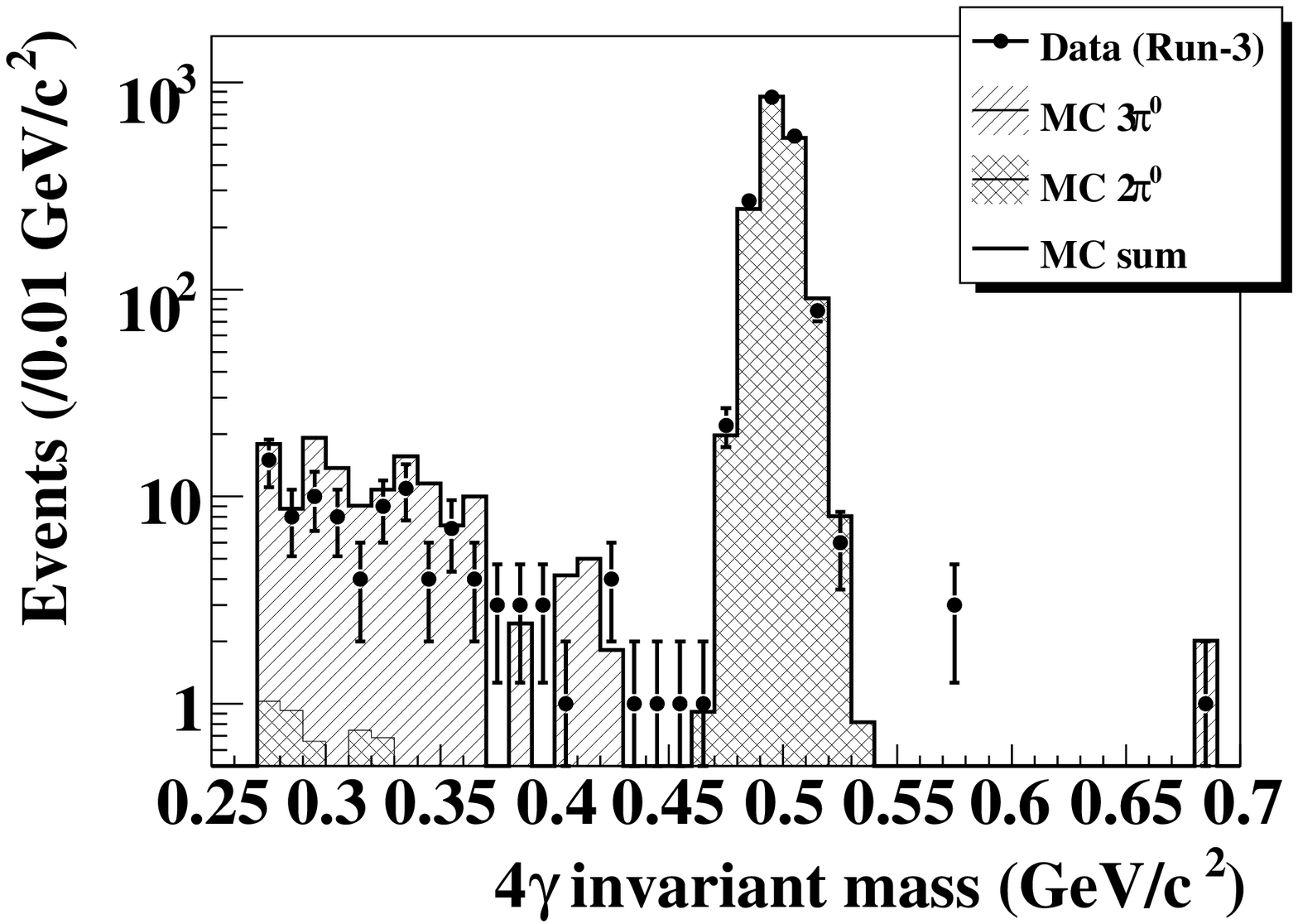}
\end{center}
\caption{Reconstructed invariant-mass distribution of events with four photons in 
the calorimeter. The points show the data and the histograms indicate the 
contribution of \kpitwos and \kpithrees decays (and their sum), as expected from 
the simulation, normalized by the number of events in the \kpitwos peak.
All veto and kinematic cuts 
except for the cut on the 4\gams invariant mass 
for the \kpitwos mode are applied.
It includes the shower-leakage correction, but does not include the correction for photo-nuclear interactions.}
\label{fig_4gmass}
\end{figure}

In addition to the distributions of kinematic variables for the \kpithrees decay mode, 
the events with four photons reconstructed in the calorimeter were analyzed
in order to verify the detection inefficiency of photon counters in the simulation.
Figure~\ref{fig_4gmass} shows the reconstructed mass distribution for four-cluster events.
In addition to the \kpitwos events at the \kls mass, there was a tail in the lower 
mass region due to contamination from \kpithree, two out of six photons of which 
escaped detection. The number of events in the tail was reproduced by our
simulations.

\begin{figure}[htbp]
\begin{center}
\includegraphics[width=.38\textwidth]{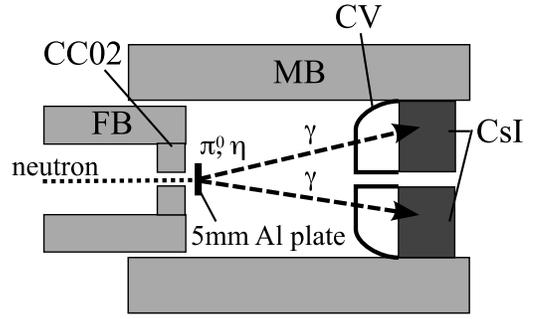}
\end{center}
\caption{
Schematic layout of the Al plate run.
}
\label{fig_pi0run_setup}
\end{figure}

\begin{figure}[htbp]
\begin{center}
\includegraphics[width=.47\textwidth]{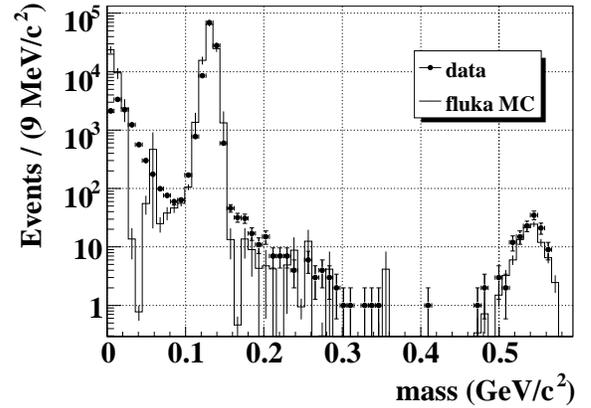}
\end{center}
\caption{Reconstructed invariant mass distribution of the two-photon event 
in the Al plate run with all the veto cuts imposed. 
The points represent data and the solid line is from FLUKA simulations. 
The peaks at 0.14 GeV/c$^{2}$ 0.55 GeV/c${^2}$ correspond 
to \piz and $\eta$ particles, respectively.}
\label{fig_alrun}
\end{figure}

A simulation with the FLUKA hadronic-interaction package to a dedicated run 
(``Al plate run")  was carried out in order to confirm the reproducibility of hadronic interactions. 
In the Al plate run, a 0.5-cm-thick aluminum plate was inserted 
into the beam at 6.5 cm downstream of the rear end of CC02, as shown in 
Fig.~\ref{fig_pi0run_setup}. Because the position of the Al plate is known, 
the invariant mass of two photons can be reconstructed from the energy and 
position of two photons. Figure~\ref{fig_alrun} demonstrates the simulation 
in which the invariant mass distribution (from \pizs mass to $\eta$ mass) 
of the events from the Al plate run was reproduced with two photons in the calorimeter.

\begin{figure}[htbp]
\begin{center}
\includegraphics[width=.47\textwidth]{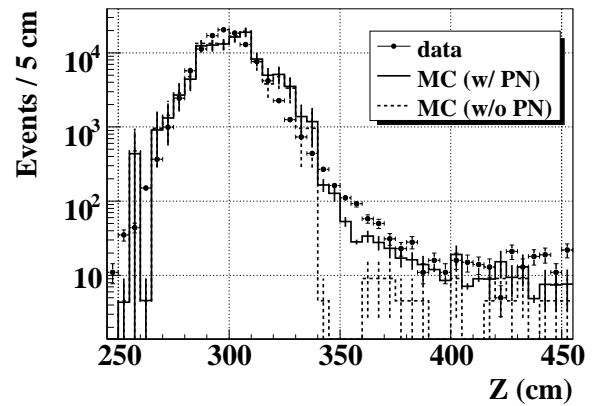}
\end{center}
\caption{Reconstructed \zvtxs distribution of two photon events in the Al plate run.
The points represent the real data and the solid (dashed) line is obtained from 
the FLUKA MC simulaiton with (without) the effects of photo-nuclear interaction.
In both cases, the shower-leakage correction was applied.
}
\label{fig_alrun_vtx}
\end{figure}

Next, by assuming the \pizs mass and reconstructing the vertex position in the Al plate run,
the effect of shower leakage and photo-nuclear interactions in the CsI was examined.
Figure~\ref{fig_alrun_vtx} shows the distribution of \zvtxs obtained from the Al plate run
by comparing the data and MC.
If the photon energy is incorrectly measured as being low due to these processes, 
the \pizs vertex position is reconstructed downstream of the true position.
Because the true \zvtxs is known in the Al plate run, we can estimate the effect 
of these processes by using the tail in the distribution of the reconstructed \zvtx.

Photo-nuclear interactions are not implemented in GEANT3, and therefore the
effects were estimated from GEANT4 simulations~\cite{geant4}. 
We estimated the probability of photo-nuclear interactions by taking the difference 
between MC samples with and without photo-nuclear interactions.
The energies of two photons were smeared according to the probability of the interaction.
As shown in Fig.~\ref{fig_alrun_vtx},
a tail on the downstream side of the peak is reproduced
by the simulations with photo-nuclear interactions.

\section{Event Selection and background estimation}

\subsection{Event selection}
Event selection consisted of veto cuts and kinematic selections.
Vetoing extra activities was the primary method used to isolate the \kznns signal events
from the background events. Kinematic selections were applied to obtain further 
rejection of background events.

\subsubsection{Veto cuts}
Particle veto with a hermetic detector system is the primary method to reject 
possible background sources. For example, events made by the \kpitwos or 
\chargedkpithrees mode can be rejected by veto cuts because these modes have 
at least two extra activities. 
The method also works for halo neutron backgrounds because the hadronic 
interactions of halo neutrons are often accompanied by extra particles 
such as protons or pions.

Energy thresholds for veto cuts were set at two levels. 
For loose veto thresholds (typically, 10 MeV), no timing cut was made 
to reduce inefficiencies caused by accidental hits before the event time. 
For tighter veto thresholds (typically, 1-2 MeV), events with a TDC value 
within roughly $\pm5\sigma$ from the on-time peak were rejected.
Combining these conditions, we balanced the efficiencies of detecting extra 
particles and the suppression of the acceptance loss caused by accidental hits.

\begin{figure}[htbp]
\begin{center}
\includegraphics[width=.4\textwidth]{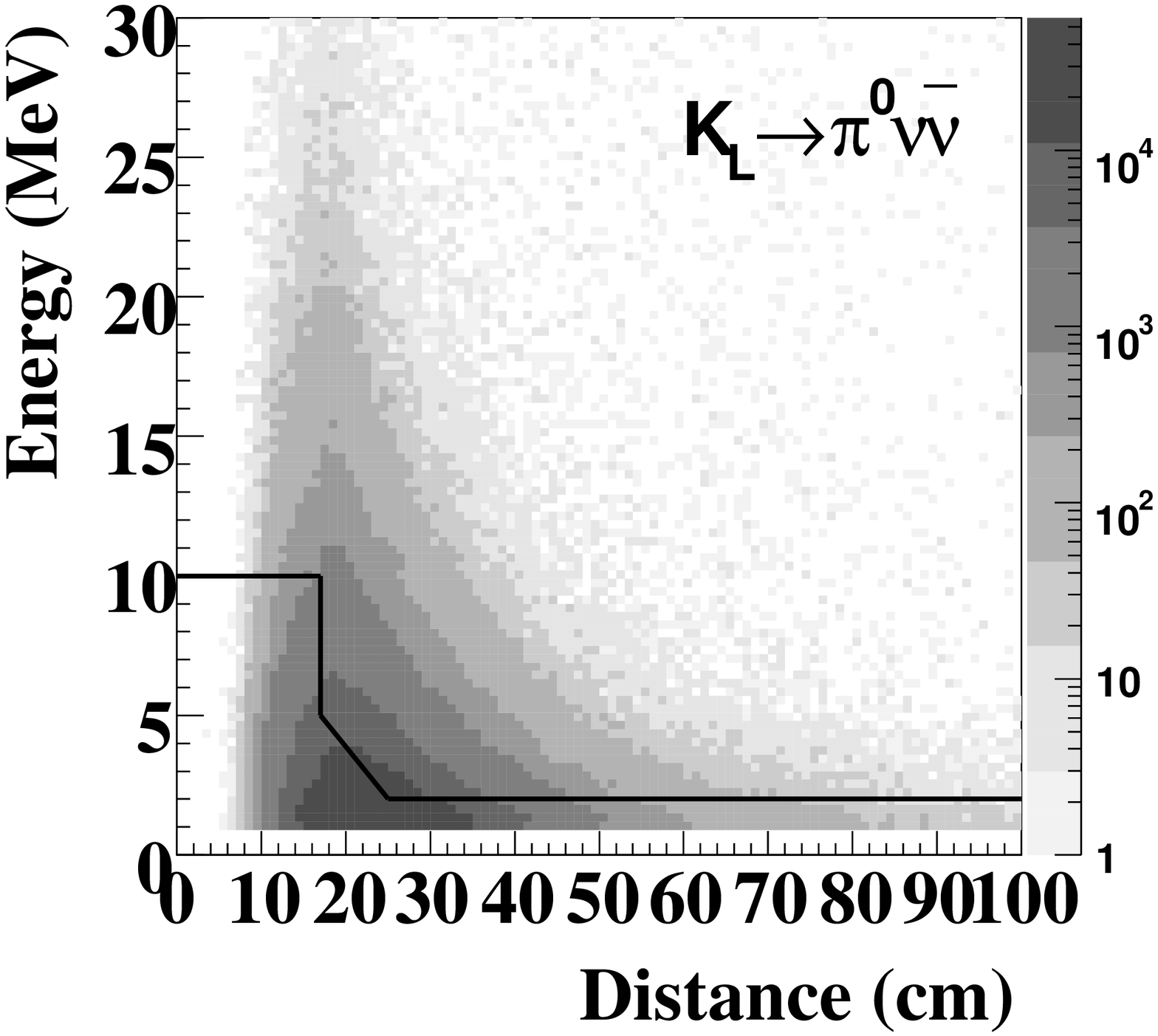}\\
\includegraphics[width=.4\textwidth]{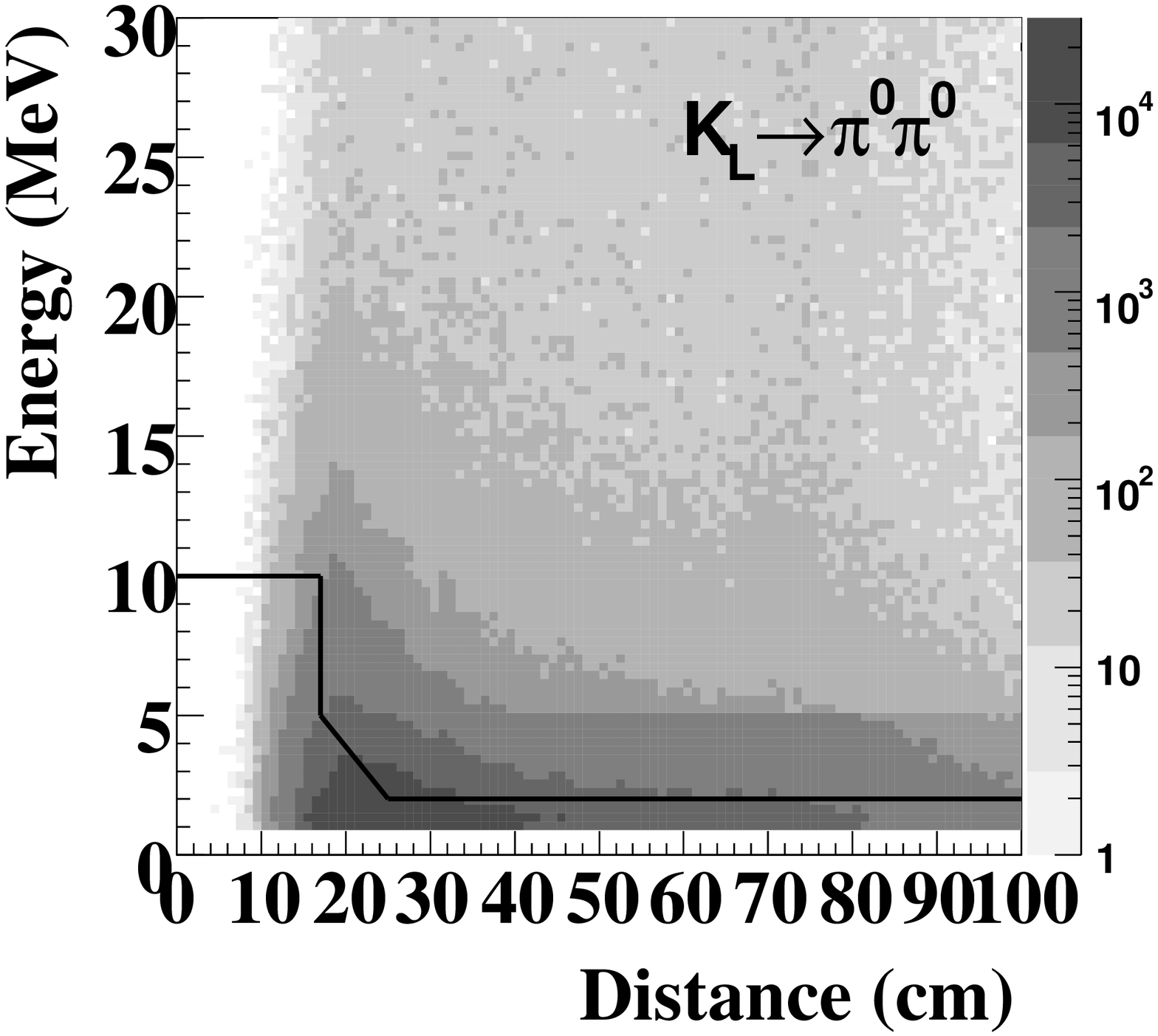}
\end{center}
\caption{
Energy deposition in a single crystal hit versus the distance
from the nearest photon cluster 
of the two-photon events from the \kznns (top) and \kpitwos (bottom) simulations.
Events above the solid line were rejected.
}
\label{fig_csiveto}
\end{figure}

\textbf{CsI veto cut:}
In addition to its main role of measuring the energy and position of photons, 
the CsI also served as a veto detector for extra photons.
Extra activities that were not reconstructed as photon clusters, {\it i.e.} 
``single crystal hits," were rejected with this cut.
However, a photon occasionally creates a single crystal hit near its genuine 
cluster due to fluctuations in the electromagnetic processes.
Thus, applying a tight energy threshold for a single crystal hit near the 
photon cluster can cause acceptance loss for signal candidates. 
To recover this loss, the energy threshold for a single crystal hit was 
determined as a function of the distance $d$ to the closest cluster 
as shown in Fig.~\ref{fig_csiveto}:
\begin{itemize}
  \item $E_{\textrm{thres.}} = 10$ MeV \;\;\; for $d < 17 $ cm, 
  \item $E_{\textrm{thres.}} = 5 - (3/8)(d-17)$ MeV for  $17 < d < 25$ cm,
  \item $E_{\textrm{thres.}} = 2$ MeV \;\;\; for $d>25$ cm,
\end{itemize}
where $E_{\textrm{thres.}}$ is the energy threshold for a single crystal hit.

\textbf{Other photon veto cuts:}
Photon veto detectors consisted of barrel counters (MB and FB), beam collar counters 
(CC00, CC02 - CC07), and a beam hole counter (BA).

Because the MB was read out at both the upstream and downstream ends, 
the energy deposit was determined as the geometrical mean of the visible energies in both sides
in order to cancel the position dependence of the light yield. 
The timing requirement for the MB was loosened when a tighter energy threshold 
was used because the timing was broadly distributed in the halo neutron backgrounds.
When a photon hit the CsI calorimeter, low-energy photons and electrons that were created 
in an electromagnetic shower occasionally went backward and hit the MB. 
This process caused a larger acceptance loss in the MB as compared to the loss
in other veto detectors.

For the collar counters in the downstream region (CC06 and CC07), 
both tighter and looser energy thresholds were
set higher than for other detectors,
in order to reduce accidental losses due to beam-induced activities.
The BA located inside the beamline rejected photons escaping into the beam hole. 
The energy thresholds for these counters are summarized in Table~\ref{table_veto}.

\textbf{Charged particle veto cuts:}
Background events produced by charged \kls decay modes were mostly rejected 
by cuts on the charged particle veto detectors, mainly CV, BCV, and BHCV.
As in the case of MB, an energy deposit in the BCV was determined as 
the geometrical mean of the hits in the upstream and downstream ends, and the timing 
requirement was loosened.
The energy thresholds for these 
counters are summarized in Table~\ref{table_veto}.

\begin{table}[htbp]
   \caption{Summary of tighter veto energy threshold for each detector.}
   \label{table_veto}
      \begin{tabular}{lc}
	\hline \hline	
	Detector & Threshold \\
	\hline \hline
	CC00 & 2 MeV\\
	\hline
	FB      & 1 MeV \footnotemark[1] \\
	\hline
	CC02 & 1 MeV \\
	\hline
	BCV   & 0.75 MeV \footnotemark[2] \\
	\hline
	MB Inner   & 1 MeV \footnotemark[2] \\
	\hline
	MB Outer   & 1 MeV \footnotemark[2] \\
	\hline
	CV Outer   & 0.3 MeV  \\
	\hline
	CV Inner   & 0.7 MeV \\
	\hline	
	CC03     & 2 MeV \\
	\hline
	CsI         & depends on $d$ \footnotemark[3] \\
	\hline
	Sandwich & 2 MeV \\
	\hline
	CC04 Scintillator & 0.7 MeV \\
	\hline
	CC04 Calorimeter & 2 MeV \\
	\hline
	CC05 Scintillator & 0.7 MeV \\
	\hline
	CC05 Calorimeter & 3 MeV \\
	\hline
	CC06 & 10 MeV \\
	\hline
	CC07 & 10 MeV \\
	\hline
	BHCV & 0.1 MeV \\
	\hline
	BA Scintillator (Run-2) & 20 MeV \footnotemark[4] \\
	BA PWO (Run-3) & 50 MeV \footnotemark[4] \\
	BA Quartz & 0.5 MIPs \footnotemark[5] \\
	\hline \hline
   \end{tabular}
   \footnotetext[1]{Sum of inner and outer layers.}
   \footnotetext[2]{Geometrical mean of upstream and downstream ends.}
   \footnotetext[3]{Detailed in the text.}
   \footnotetext[4]{Summed over layers.}
   \footnotetext[5]{Determined by AND logic of scintillator/PWO and quartz.}
\end{table}

\subsubsection{Kinematic selections}
Kinematic selections were applied to discriminate the signal from background events 
by using the information of two photons in the CsI calorimeter. 
These selections can be categorized into three types: photon-cluster quality 
selections, 
selection on photons, and \pizs selections.

\textbf{Photon-cluster quality selections:}
Photon-cluster quality selections were developed from the information of each photon cluster.
The energy was required to be 
$E_H > 250$ MeV and $E_L > 150$ MeV, where $E_H$ and $E_L$ are
the energies of the higher- and lower-energy photons, respectively.
We also made selections on the incident position, size, profile of the energy distribution,
and timing dispersion of each photon cluster.
These cuts were applied to ensure that two photon clusters were from the \kznns decay
and were distinguished from fake clusters created by hadronic showers or photons produced by
halo neutron interactions.

\begin{figure}[htbp]
\begin{center}
\includegraphics[width=.47\textwidth]{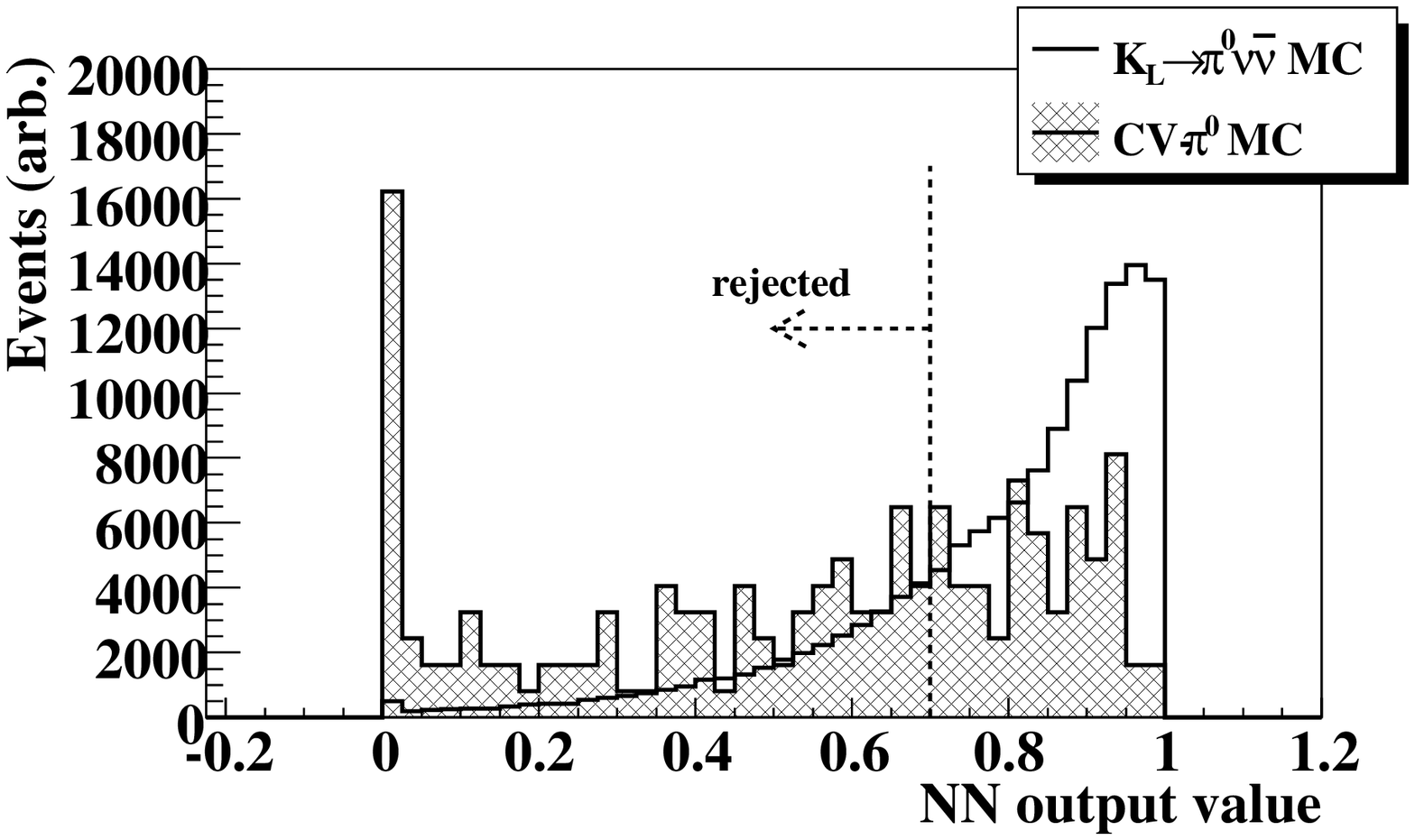}\\
\includegraphics[width=.47\textwidth]{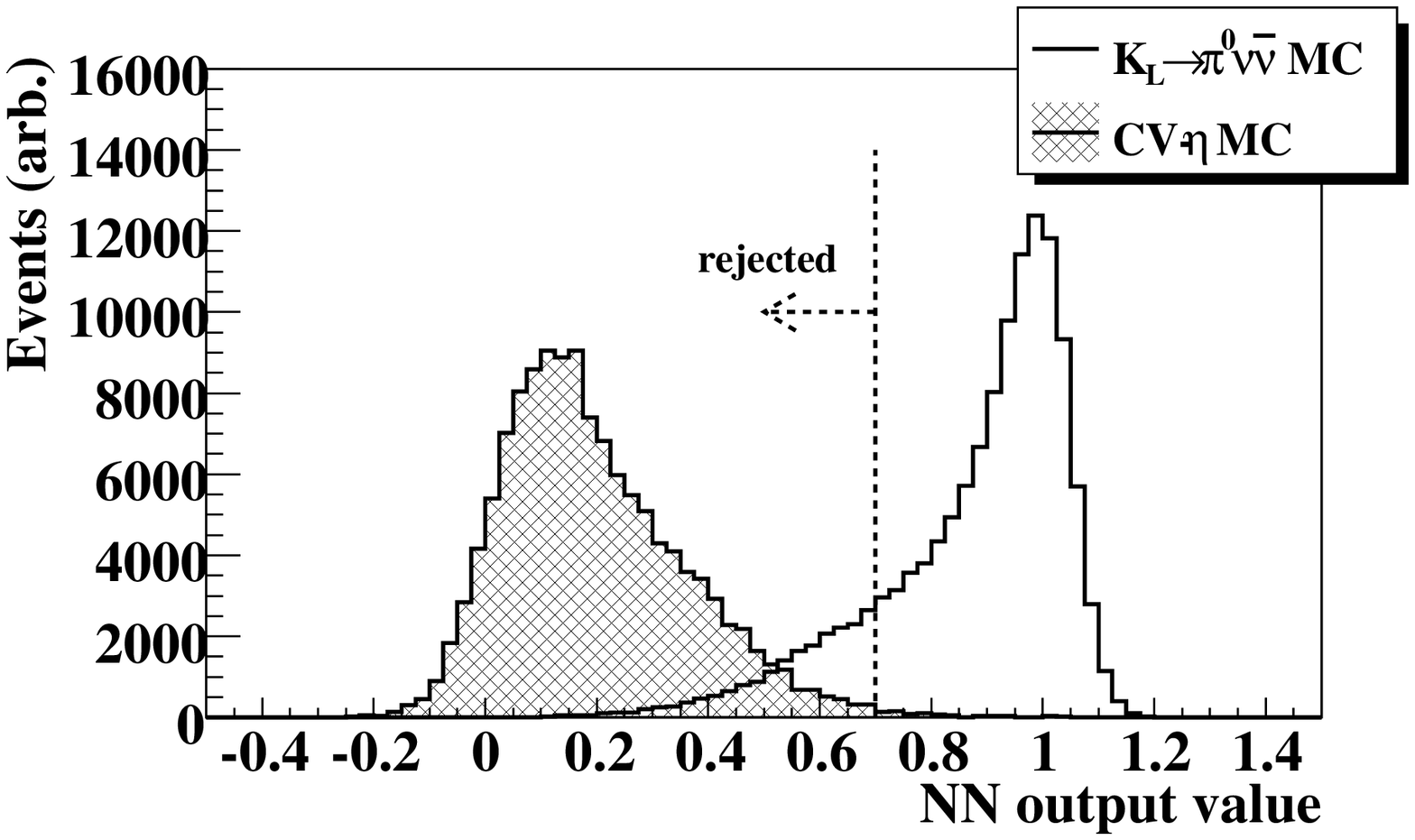}
\end{center}
\caption{
Distributions of the output values obtained by two neural network selections 
for the signal mode (hollow) and the background events (hatched).
The top figure shows the fusion neural network selection, in which the background 
is CV-\piz. The bottom figure shows the CV-$\eta$ neural network selection, 
in which the background is CV-$\eta$. In both plots, all analysis cuts except 
for the respective neural network cut are imposed for the signal mode. 
Several cuts are loosened for the background case in order to enhance the events.
}
\label{fig_nncuts}
\end{figure}

In addition, two neural network selections were developed pertaining to the shape 
of a photon cluster. The first neural network was used to reduce events with 
clusters that overlapped those from other photons or associated particles and reconstructed 
as a single cluster (``fusion cluster"). In the case of \kpitwos background, a fusion cluster 
results in an inefficiency to extra photons; in the case of CV-\pizs background, 
it results in a larger photon energy such that the event vertex 
comes into the signal region. This neural network was trained by single 
and fusion clusters obtained from the \kpitwos MC samples. 
The second neural network selection was used to reject the CV-$\eta$ background.
Because $\eta$ particles were produced near the front end of CsI and had a larger invariant 
mass than a \piz, the two photons produced by $\eta$ decay tended to 
have large incident angles to the CsI calorimeter.
This neural network was trained by the MC samples of \kznns and CV-$\eta$ background. 
The rejection power of these two neural network selections is
shown in Fig.~\ref{fig_nncuts}, with the distribution of the output values from 
the neural network for the signal and background modes.

\textbf{Selection on photons:}
We required the distance between two photons to be greater 
than 15 cm, and the timing difference between them to be 
$-9.6 < T(E_H) - T(E_L) < 18.4$ ns, where $T(E_H)$ and $T(E_L)$ are the 
timing of the higher and lower energy photons, respectively.
In addition, the energy balance of two photons defined as 
$(E_{H}-E_{L})/(E_{H}+E_{L})$ , was required to be less than 0.75.
These selections were needed mainly to suppress accidental activities.

\textbf{\pizs selections:}
By using the information for reconstructed \piz's, several selection criteria were imposed. 
We required the kinetic energy of the reconstructed \piz's to be less 
than 2 GeV to suppress neutron backgrounds with high energy.
In addition, the reconstructed \piz's had to be kinematically consistent 
with a \kznns decay within the proper \kls momentum range.
A neural-network selection was applied to the discrepancy between two angles.
One angle was calculated by connecting the incident position of photons and the position of reconstructed \piz's,
and the other angle was estimated by the neural network based on the cluster shape.
This selection suppressed CV-\pizs and CV-$\eta$ backgrounds 
because the discrepancy of the angle increased in these processes.
To reduce the \kggs background, we calculated the opening angle between 
the two photon directions projected onto the CsI calorimeter plane, 
and required it to be less than 135 degrees (``acoplanarity angle cut").

\subsubsection{Signal region}
We set the signal region in the \zvtx-\pts plane to be 
$340 \leq Z_{VTX} \leq 500 $ cm and $0.12 \leq P_T \leq 0.24$ GeV/$c$.
The requirement on \zvtxs was determined to avoid background \piz's 
coming from the CC02 (upstream) and CV (downstream).
The lower limit for \pts was set to reduce the \kpitwos and CV-$\eta$ 
backgrounds; these events cluster in the low \pts region due to kinematics 
and a large halo neutron flux near the beam center, respectively. 
The upper bound for \pts was determined by the kinematic limit of the 
\kznns decay, whose maximum value is 0.231 GeV/$c$.

\subsection{Background estimation}

\subsubsection{Halo neutron background}
The halo neutron background was estimated from combined MC simulations 
with FLUKA and GEANT3.

\begin{figure}[htbp]
\begin{center}
\includegraphics[width=.47\textwidth]{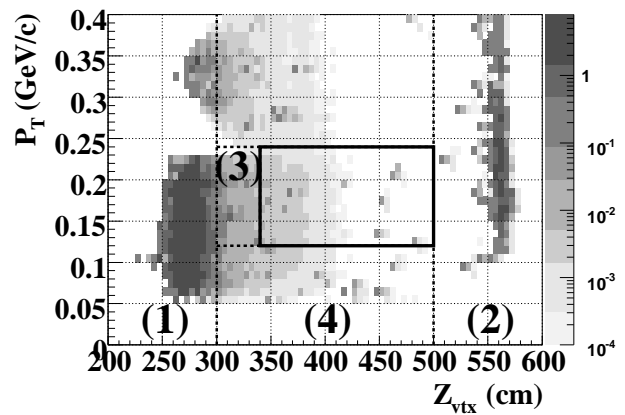}
\end{center}
\caption{
Scatter plot of $P_{T}$ vs.\ the reconstructed $Z$ position after applying 
all cuts to the MC samples for background simulation, including halo neutron and \kpitwos 
backgrounds. The region bounded by the solid line shows the signal box.
Events around $Z_{VTX} = 275$ cm are halo neutron events reconstructed at 
the position of CC02, and those around $Z_{VTX} = 560$ cm are events 
reconstructed at the CV. The numbers of events in each region are listed 
in Table~\ref{table_outside}.
}
\label{fig_bgestim}
\end{figure}

Figure~\ref{fig_bgestim} shows the distribution of background events in the 
\zvtx-\pts plane as estimated from the MC samples.
The outside of the signal box was divided into four regions, 
as shown in Fig.~\ref{fig_bgestim}. 
The number of events in each region was dominated by the halo neutron backgrounds.
The numbers of halo neutron backgrounds were normalized 
to the number of events in the upstream region (shown as Region-(1)). 
The numbers for events outside the signal box 
were compared between the data and the MC, as listed in Table~\ref{table_outside}.
The numbers of events in each region are consistent within the statistical uncertainties
between the data and the MC.
Possible impacts of discrepancies outside the signal box on the estimation inside the signal box
are included in the systematic uncertainties below, and will be discussed in detail 
in a later section (Sec.\ VI.B.).

For estimating the CC02-\pizs background, the effects of shower leakage and photo-nuclear
interactions described in Sec.\ III.D were taken into account.
After applying all event selections, we estimated $0.66 \pm 0.33_{\textrm{stat}} 
\pm 0.20_{\textrm{syst}}$ background events in the signal box for the combined 
data sample of Run-2 and Run-3. 

For CV-\pizs events, a fusion cluster or a cluster made by a neutron produces 
background. After applying all event selections, no events inside 
the signal box were obtained with 3 times larger statistics as compared to 
the combined data sample of Run-2 and Run-3. An upper limit of 0.36 events 
at the $1 \sigma$ level was set for this process.

CV-$\eta$ events were strongly suppressed by the dedicated neural-network cut 
on the cluster shape. A value of $0.19 \pm 0.08_{\textrm{stat}} \pm 
0.10_{\textrm{syst}}$ events was estimated for this process.

\begin{table}[htbp]
\caption{
Estimated numbers of events outside the signal region, compared with the
combined data samples of Run-2 and Run-3.
Note that the numbers of events estimated by the MC simulation were normalized in Region-(1).
Errors in the MC estimates indicate statistical uncertainties.
}
\label{table_outside}
\begin{tabular}{rlcc}
\hline \hline
\multicolumn{2}{c}{Region} & Data & MC estimation \\
\hline
Region-(1) & (CC02)       & 360 & $360.0 \pm 15.6$ \\
Region-(2) & (CV)            & 101 & $77.2 \pm 5.6$ \\
Region-(3) & (upstream)  & 8     &  $5.9 \pm 1.1$  \\
Region-(4) &  (low-\pt)     & 8      & $2.9 \pm 0.9$   \\
\multicolumn{2}{c}{Signal box} & (masked) & $0.87 \pm 0.34$ \\
\hline\hline
\end{tabular}
\end{table}

\subsubsection{\kls background}
The \kls background was estimated by using GEANT3-based MC simulations 
that were performed separately for Run-2 and Run-3 conditions.
Among the backgrounds from \kls decay, the \kpitwos mode made the 
largest contribution. 
The amount of Monte Carlo statistics for the \kpitwos mode was roughly 70 (60) times 
that for the real data of the Run-2 (Run-3) sample.
In Run-2, two events remained after applying all event selections in 
the simulations, and in Run-3, no events remained. 
The two remaining events both corresponded to the case that 
two photons in the CsI came from the same \pizs.
One extra photon with high energy ($\sim$ 1 GeV) went through the CsI (``punch-through"), 
and the other extra photon with low energy ($\sim$ 10 MeV) hit the MB and failed 
to be detected by the fluctuation in an electromagnetic shower.
The estimated number of background events from \kpitwos decay was 
estimated as $(2.4 \pm 1.8_{\textrm{stat}} \pm 0.2_{\textrm{syst}}) 
\times 10^{-2}$ for the combined data sample of Run-2 and Run-3.

For the MC \kggs samples whose statistics corresponded 
to 3.8 and 8.7 times the Run-2 and Run-3 data, respectively,
all of the event selection cuts except for the acoplanarity angle cut were 
imposed and no events remained in the signal region.
Furthermore, the acoplanarity angle cut strongly suppressed the \kggs events, 
with a typical rejection of $1\times 10^5$.
We concluded that the background due to \kggs was negligible in this analysis.

For charged modes, the branching ratios of the \chargedkpithrees and 
\semileptonics $(l=e, \mu)$ modes are too large to generate sufficient statistics 
in the simulations.
The background from these modes 
were estimated by assuming that the inefficiency of charged-particle vetos was 
$1.0\times10^{-4}$ for the CV, and $1.0\times10^{-3}$ for BCV and BHCV \cite{chineff}. 
By applying the event weight calculated from the inefficiency, the numbers of 
background events from these modes were estimated to be 
$4.2 \times 10^{-4}$ for \kpies and less than $1.0 \times 10^{-4}$  for \chargedkpithree.
Here, the \kpies mode had the largest contribution to the background events among 
\semileptonics modes because a charge-exchange interaction of a \pims ($\pi^- + p 
\rightarrow \pi^0 + n$) and annihilation of $e^+$ prevent the detection of 
charged particles.

\subsubsection{Other background sources}
In addition to the backgrounds from halo neutron and \kls decay, two other 
possible background sources were considered. The first one is the 
{\it backward-going \pizs background}. 
When the halo neutron interacts in the end cap of the vacuum vessel located 
2 m downstream of the CsI, \piz's are occasionally produced in the upstream direction. 
Because we were unable to discriminate whether photons came from the front or the back, 
these events were reconstructed as coming from the front and ended up inside the 
signal box.
We estimated this background by using FLUKA simulations with 20 times the amount 
of statistics as compared to the combined data sample of Run-2 and Run-3.
No events remained after applying all cuts, and the number of background events 
was determined to be less than 0.05 events.

The second additional background source is the {\it residual gas background} 
that occurs from interactions of beam neutrons with the residual gas.
This process is well suppressed by a high-vacuum system that provided 
a very low pressure of $10^{-5}$ Pa.
To estimate the background, we carried out a dedicated run at atmospheric 
pressure, accumulating statistics that were roughly equivalent to 0.6\% of the 
combined data sample of Run-2 and Run-3. We obtained 6867 candidate events 
with loose event selections. By considering a reduction factor of $10^{-10}$ 
due to the air pressure, this background was concluded to be negligible.

We investigated possible contributions from unknown background sources by
loosening the selection criteria for the data and MC. 
The numbers of events in the data sample with loosened cuts were consistent 
with the prediction with the known background sources.

In total, the estimated number of background events was $0.87 \pm 0.41$, 
where the CV-\pizs and the backward-going \pizs backgrounds were not included.
Table~\ref{table_bgestim} summarizes estimates for the numbers of background 
events.

\begin{table}
\caption{
Estimated number of background events.
}
\label{table_bgestim}
\begin{tabular}{ccc}
\hline \hline
\multicolumn{2}{c}{Background source}          & Estimated number of BG\\
\hline
halo neutron BG  & CC02-\piz       & $0.66 \pm 0.39$ \\
 & CV-\piz       & $<0.36$ \\
 & CV-$\eta$       & $0.19 \pm 0.13$ \\
\hline
\kls decay BG  & \kpitwo       & $(2.4 \pm 1.8)\times 10^{-2}$ \\
                           & \kggs       & negligible \\
                           & charged modes       & negligible (${\cal O}(10^{-4})$) \\
\hline
other BG           & backward \piz       & $< 0.05$ \\
                          & residual gas       & negligible (${\cal O}(10^{-4})$)\\  
\hline
\multicolumn{2}{c}{total}          & $0.87 \pm 0.41$ \\
\hline \hline
\end{tabular}
\end{table}

\section{Sensitivity and results}

\subsection{Principle of normalization}

The single event sensitivity (S.E.S.) is represented as 
\begin{equation}
\textrm{S.E.S.} (K_L^0\rightarrow \pi^0\nu\bar{\nu}) = 
\frac{1}{N(K_L^0\:\textrm{decays}) \times A_{\textrm{signal}}} \:,
\label{eq_SES}
\end{equation}
where $N(K_{L}^{0}\:\textrm{decays})$ denotes the number of \kl's that decayed 
in the fiducial region and $A_{\textrm{signal}}$ denotes the acceptance for the signal mode.
The value of $N(K_{L}^{0}\:\textrm{decays})$ is determined from the normalization mode as
\begin{equation}
 N(K_L^0\:\textrm{decays}) = \frac{N_{\textrm{norm.}}^{\textrm{data}}}{A_{\textrm{norm.}}
\times BR_{\textrm{norm.}}}   \;,
\label{eq_nkl}
\end{equation}
where $N_{\textrm{norm.}}^{\textrm{data}}$ is the number of observed events from 
the normalization mode and $A_{\text{norm.}}$ and $BR_{\textrm{norm.}}$ represent 
the acceptance and branching ratio of the mode, respectively.
Substituting Eq.~\ref{eq_nkl} into Eq.~\ref{eq_SES}, the S.E.S. is obtained 
from the number of remaining events in the normalization mode and the ratio of
acceptances between the signal and the normalization modes.
By taking this ratio, uncertainties arising from variations in beam condition, 
{\em etc.}, can be canceled.

We examined three decay modes, namely, \kpithree, \kpitwo, and \kggs, to obtain 
the number of \kls decays, because they were fully reconstructed and clearly identified.
Because these three decay modes have different numbers of photons in the final states, 
we could cross-check the reliability of the clustering method. 
The mean value of the accepted \kls momentum was also different, which provided a 
check in the full range of the accepted \kls momentum for the \kznns mode.

\subsection{Analysis of normalization modes}
\subsubsection{\kls reconstruction}
For the \kpithrees and \kpitwos modes, \kl's were reconstructed from 
the six and four photons in the CsI calorimeter, respectively.
In the reconstruction, the number of possibile combinations of pairs of 
the two photons was $(_6C_4 \times _4C_2 \times _2C_2)/3! = 15$ for 
\kpithrees and $(_4C_2 \times _2C_2)/2! = 3$ for \kpitwo. 
The decay vertex location can be calculated from the energy and position 
of the two photons of each pair by assuming the \pizs mass.
To find the best combination of the photons, the variance in the 
reconstructed vertex points was calculated, named ``pairing $\chi_Z^2$," 
for all the possible combinations.
The pairing $\chi_Z^2$ was defined as
\begin{equation}
\chi_Z^2 = \sum_{i=1}^n \frac{  ( Z_i - \bar{Z} )^2}{ \sigma_i^2} \:,
\label{eq_pairing_chi}
\end{equation}
where $i$ runs over the two-photon pairs reconstructing \piz, 
({\it e.g.}, $n=3$ for \kpithrees and $n=2$ for \kpitwo.), 
$Z_i$ is the vertex point of $i$-th two-photon pair, $\sigma_i$ is the 
resolution in reconstructing $Z_i$ calculated from the energy and position 
resolutions of the two photons, and 
\begin{equation}
\bar{Z} =  \frac{ \sum_{i=1}^n  Z_i/\sigma_i^2}{ \sum_{i=1}^n 1/\sigma_i^2} \:.
\label{eq_barZ}
\end{equation}
The decay vertex of the \kls was determined as $\bar{Z}$ for the combination 
with minimum $\chi_Z^2$.
We required the minimum pairing $\chi_Z^2$ to be less than 3.0 and the difference 
between the next-to-minimum one to be greater than 4.0, in order to reduce 
incorrect paring.

For the \kggs mode, \kl's were reconstructed from the two photons in the 
CsI calorimeter by assuming the \kls mass.

Several analysis cuts were imposed on the events in each decay mode to remove 
contaminations of other decay modes. 
Photon veto cuts were important to detect the modes with extra photons,
such as \kpithrees to the \kpitwos events, and \kpitwos to the \kggs events.
The energy threshold for the veto in each subsystem was the same as that 
used in the \kznns analysis (see Table~\ref{table_veto}).
The reconstructed mass of six photons in the \kpithrees events and four photons 
in the \kpitwos events had to be consistent with the \kls mass (from 0.481 to 
0.513 GeV/$c^2$). 
The decay vertex point of \kls also had to be located in the fiducial region 
(from 340 to 500 cm), as in the case of \kznn.
Some additional cuts, such as the neural network fusion cluster selection, were also
imposed on these decay modes.

\subsubsection{Number of \kls decays}
The acceptance of each mode was estimated from Monte-Carlo simulations.
Because, in the simulations, \kl's were generated at the exit of the 
last collimator (C6), the probability of \kls decay in the fiducial region, 
$(2.14 \pm 0.02)$\%, was calculated separately
and was taken into account in calculating the number of \kls decays.
Losses due to accidental activities in the detector 
were also included in the acceptance.

Table~\ref{table_flux} shows a summary of the estimated numbers of \kls decays 
from the three decay modes \kpithree, \kpitwo, and \kgg.
The difference among the three modes was within the systematic uncertainties,
and considered to come from the CsI veto, because it has the largest systematic uncertainties
as well as a dependence on the number of photons.
We adopted the number obtained from \kpitwo,
$(8.70 \pm 0.17_{\textrm{stat}} \pm 0.59_{\textrm{syst}}) \times 10^9$ for the combined 
data sample of Run-2 and Run-3, as the normalization of this analysis 
because the energy distribution of photons in the CsI calorimeter from \kpitwos
was similar to that expected to \kznn.
Estimates of the systematic uncertainties are described later.

\begin{table*}[htbp]
\caption{
Estimated numbers of \kls decays calculated from the three decay modes in the 
combined sample of Run-2 and Run-3 data.
Uncertainties in the acceptances are statistical ones due to the amount of the MC samples.
$N_{\textrm{norm}}^{\textrm{data}}$ is the number of events obtained from the 
real data after imposing all the analysis cuts. 
In the \kpitwos mode, $N_{norm}^{data}$ is obtained by subtracting the 
contamination from the \kpithrees mode.
Statistical uncertainties in the number of \kls decays include an ambiguity 
in $N_{\textrm{norm}}^{\textrm{data}}$, and systematic ones include an 
ambiguity from the reproducibility of the MC (described later) and
statistical uncertainties of the MC in the acceptance estimate.
}
\label{table_flux}
\begin{tabular}{cccc}
\hline
mode & acceptance & $N_{norm}^{data}$ & $N(K_L^0 \mathrm{decays})$ \\
\hline
\kpithrees $\;$& $\;$ $ (7.21 \pm 0.06) \times 10^{-5} $ $\;$ & $\;$ 118334 $\;$ 
  & $\;$ $ (8.41 \pm 0.03_{\textrm{stat}} \pm 0.53_{\textrm{syst}}) \times 10^{9} $ \\
\kpitwos  $\;$  & $\;$ $ (3.42 \pm 0.03) \times 10^{-4} $ $\;$ & $\;$ 2573.9 $\;$ 
  & $\;$ $ (8.70 \pm 0.17_{\textrm{stat}} \pm 0.59_{\textrm{syst}}) \times 10^{9} $ \\
\kggs        $\;$ & $\;$ $ (7.18 \pm 0.03) \times 10^{-3} $ $\;$ & $\;$ 35367 $\;$ 
  & $\;$ $ (9.02 \pm 0.05_{\textrm{stat}} \pm 0.51_{\textrm{syst}}) \times 10^{9} $ \\
\hline
\end{tabular}
\end{table*}

\subsection{Acceptance and single event sensitivity}
\subsubsection{Signal acceptance}
The acceptance for the \kznns decay was estimated from Monte-Carlo simulations.
The raw acceptance was calculated by dividing the number of remaining events
after imposing all the analysis cuts by the number of \kznns decays generated in 
the simulation.
It was 1.40\% for the Run-2 and 1.39\% for the Run-3 data, respectively.

In the raw acceptance, the losses caused by geometrical acceptance, veto cuts, 
kinematic selections, and selection on fiducial region were included. 
The geometrical acceptance to detect two photons in the CsI was approximately 20\%.
The loss by veto cuts, or self-vetoing, was about 50\%, which was dominated by the CsI and MB vetoes.
The loss in the CsI veto was caused by hits accompanied by the genuine photon cluster, as shown in 
Fig.~\ref{fig_csiveto}, and the loss in the MB was caused by photons escaping from the
front face of the CsI and hitting the MB.
The acceptance of the kinematic and signal region selections was approximately 15\%.
These values were estimated through MC simulations of the \kznns mode.
The evaluation of the acceptance loss was supported by the fact that
the acceptance losses in the normalization modes (\kpithree, \kpitwo, and \kgg)
were reproduced by the simulation.

\begin{figure}[htbp]
\begin{center}
\includegraphics[width=.47\textwidth]{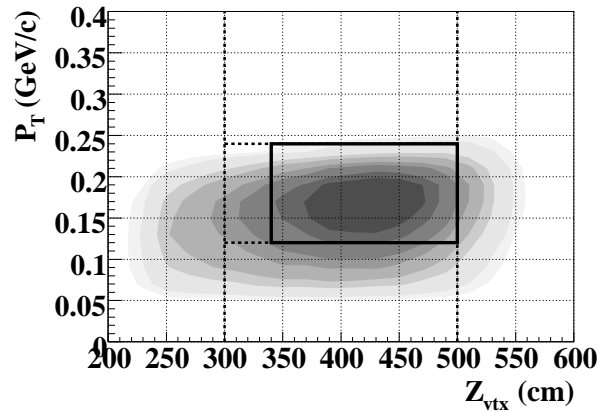}
\end{center}
\caption{
Density plot of \pts vs. the reconstructed $Z$ position for the \kznns Monte Carlo events
after imposing all of the analysis cuts.
The box indicates the signal region for \kznn.}
\label{fig_ptz_signal}
\end{figure}

The acceptance loss due to accidental activities in the detector 
was estimated from real data taken with the TM trigger. 
The accidental loss was estimated to be 20.6\% for the Run-3 data,
in which the losses in MB (7.4\%) and BA (6.4\%) were major contributions.
For Run-2, the accidental loss was estimated to be 17.4\%; the difference 
between Run-2 and Run-3 was due to the difference in the BA counters used 
in the data taking. 
The acceptance loss caused by 
the selections on the timing dispersion of each photon cluster and 
on the timing difference between two photons
was estimated separately by using real data and was obtained to be 8.9\%. 
Thus, the total acceptance for the \kznns was $(1.06\pm0.08)$\% for Run-2 and
$(1.00\pm0.06)$\% for Run-3 case, where the errors are dominated by the 
systematic uncertainties that are discussed later. Figure~\ref{fig_ptz_signal} 
shows the distribution of the MC \kznns events in the scatter plot of 
\pt-\zvtxs after imposing all of the other cuts.

\subsubsection{Single event sensitivity}
By using the number of \kls decays and the total acceptance, 
the single event sensitivity for \kznns was
$(1.84 \pm 0.05_{\textrm{stat}} \pm 0.19_{\textrm{syst}} )\times 10^{-8}$ for Run-2,
$(2.80 \pm 0.09_{\textrm{stat}} \pm 0.23_{\textrm{syst}} )\times 10^{-8}$ for Run-3,
and
$(1.11 \pm 0.02_{\textrm{stat}} \pm 0.10_{\textrm{syst}} )\times 10^{-8}$ in total.

\subsection{Results}
\begin{figure}[htbp]
\begin{center}
\includegraphics[width=.47\textwidth]{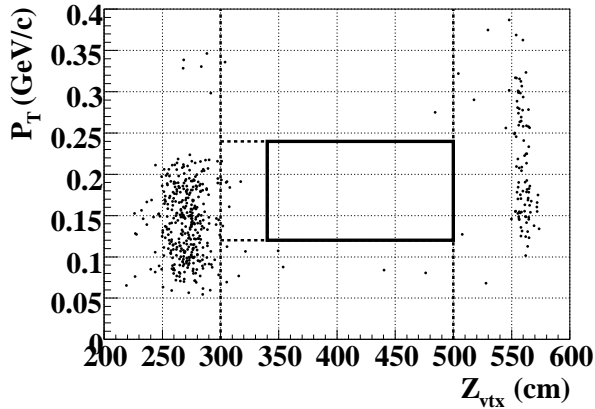}
\end{center}
\caption{Scatter plot of \pts vs.\ the reconstructed $Z$ position for the events 
with all of the selection cuts imposed. The box indicates the signal region for \kznn.}
\label{fig_final}
\end{figure}

After finalizing all of the event selection cuts, the candidate events inside the 
signal region were examined. No events were observed in the signal region, as shown 
in Fig.~\ref{fig_final}. 
An upper limit for the \kznns branching ratio was set to be $2.6\times10^{-8}$ at 
the 90\% confidence level, based on Poisson statistics. 
The result improves the limit previously published~\cite{run2}
by a factor of 2.6.

\section{Systematic uncertainties}
Although the systematic uncertainties were not taken into account 
in setting the current upper limit on the branching ratio,
we will describe our treatment for them 
to provide a thorough understanding of the experiment. 
In particular, systematic uncertainties of the single event sensitivity and 
background estimates due to halo neutrons are discussed in order.

\subsection{Uncertainty of the single event sensitivity}
The systematic uncertainty of the single event sensitivity was evaluated 
by summing the uncertainties of the number of \kls decays 
and the acceptance of the \kznns decay. 
Because the calculation of the former also includes 
the acceptance of the normalization modes, 
the acceptance of both the normalization mode and the \kznns mode are relevant to the acceptance evaluation
by the Monte Carlo simulations.
To estimate the uncertainties in the acceptance calculation,
we utilized the fractional difference between data and the simulation in each selection criterion,
defined by the equation
\begin{equation}
F^i = \frac{A_{\textrm{data}}^i - A_{\textrm{MC}}^i}{A_{\textrm{data}}^i} 
\end{equation}
where $A_{\textrm{data}}^i$ and $A_{\textrm{MC}^i}$ denote the acceptance values of the $i$-th cut, 
calculated as the ratio of numbers of events with and without the cut,
for the data and MC simulations, respectively. 
In $F^i$, the acceptance was calculated with all the other cuts imposed.
The systematic uncertainty of the acceptance was evaluated
by summing all the fractional differences in quadrature, 
weighted by the effectiveness of each cut, as 
\begin{equation}
\sigma_{\textrm{syst.}}^2 = \frac{\sum_{i=\textrm{cuts}} 
(F^i/A_{\textrm{data}}^i)^2}{\sum_{i=\textrm{cuts}}(1/A_{\textrm{data}}^i)^2} \:.
\end{equation}

For the three decay modes used in the normalization, 
\kpithree, \kpitwo, and \kgg, 
the calculated uncertainties were 5.2\%, 5.7\%, and 3.6\% (in Run-3), respectively.
The acceptances of the CsI veto cut had the largest uncertainties in all decay modes.
The number of \kls decays was obtained by using the \kpitwos mode, 
and thus its uncertainty is quoted as the same value for the \kpitwos mode.

For the acceptance of the \kznns mode, the same systematic uncertainties as the 
\kpitwos mode were adopted because there were no signal candidates in the data 
to be compared with the MC simulations.

The systematic error of the single event sensitivity was evaluated to be a
quadratic sum of the uncertainties of the number of \kls decays and the 
acceptance of the \kznns decay.  It was 10.3\% in Run-2 and 8.2\% in Run-3, 
respectively.

\subsection{Uncertainty of the halo neutron backgrounds}
The systematic errors of halo neutron backgrounds were also estimated 
by utilizing fractional differences between data and the simulations.
There were two kinds of cuts in our analysis, 
veto cuts and kinematic selections.
For the former, the method was unable to be used directly because 
veto cuts had been applied in the early stages of the MC simulations  
to save the computing time. 
Thus, for the systematic uncertainty due to veto cuts, 
the same value obtained in the \kpitwos analysis was assigned.
For the kinematic selections, the same method as described in 
the previous section was used.
The acceptance of each cut was calculated 
with all veto cuts imposed except kinematic selections.
In total, the systematic uncertainties due to fractional differences 
were calculated to be 31\%, 32\%, and 44\% 
for CC02-\piz, CV-\piz, and CV-$\eta$ backgrounds, respectively.

In addition, the uncertainties in the normalizations of the MC 
simulations were taken into account. 
Because the normalization was determined by 
using the number of events in the CC02 region (Region-(1)),
there can be ambiguity in estimating the CV-related backgrounds. 
As shown in Table~\ref{table_outside}, 
there was a 24\% difference between data and MC simulation 
in the downstream region (Region-(2)).
Even though they were statistically consistent, we assigned the 
difference as an additional systematic uncertainty of CV-\piz.
For the CV-$\eta$ case, further ambiguity due to the reproducibility 
of $\eta$ production should be considered.
It was estimated to be 24\% from the difference between data and MC
simulations in the numbers of $\eta$ events in the Al plate run, 
as shown in Fig.~\ref{fig_alrun}.

The total systematic uncertainties were calculated by summing up
the contributions quadratically, 
to be 31\%, 40\%, and 55\% for CC02-\piz, CV-\piz, and CV-$\eta$
backgrounds, respectively.

\section{Conclusion and discussion}
The E391a experiment at the KEK-PS was the first dedicated experiment 
for the \kznns decay. Combining the periods of Run-2 and Run-3, 
the single event sensitivity reached $1.11\times10^{-8}$.
No events were observed inside the signal region and 
the new upper limit on the branching ratio of the \kznns decay was set 
to be BR $< 2.6 \times 10^{-8}$ at the 90\% confidence level.
The result improves the previous published limit given by the Run-2 
analysis by a factor of 2.6, 
and the E391a experiment as a whole has improved the limit 
from previous experiments by a factor of 20. 

The E391a experiment was also the first step of our step-by-step approach
toward the accurate measurement of the \kznns decay. The experiment was planned 
to confirm our experimental approach, and this purpose was well achieved. 
Several points need to be noted.

First, we found solutions to several technical issues, 
such as the pencil beamline, differential pumping for ultra-high vacuum, 
low-threshold particle detection with a hermetic configuration, in situ calibration, {\it etc.},
which can be successively used in the next step. 
We encountered some technical problems that exist in the current apparatus, 
such as insufficient thickness and segmentation of the CsI calorimeter, 
the structure of the CV, the limitation of the BA in an environment with higher counting rate, {\it etc.}.
They will be improved in the next experiment.
  
Secondly, we were able to control the systematic uncertainties to be small in the estimate 
of the single event sensitivity, as described in the previous section. 
A small systematic error is essential for an accurate determination of the 
branching ratio. We are confident that the branching ratio of the decay \kznns 
can be measured accurately with this method. 

A third point concerns the understanding and estimation of backgrounds. 
In the experiment for the decay \kznn, elimination of all possible backgrounds 
is the only effective way to identify the decay, 
and it can be achieved by a profound understanding of backgrounds.

The dominant source of backgrounds was the result of beam interactions.  
Although the background from this source was considered to be as 
serious as those from \kls decays in the experiment, 
its understanding and estimation were very difficult to assess before the work 
reported here.  The background mechanisms were clearly understood and divided 
into three different sources, CC02, CV-$\pi^0$, and CV-$\eta$. 
Methods were developed to estimate them with rather small systematic errors. 
Based on this experience, we clearly know the direction of upgrades to minimize 
the beam backgrounds in the next step. 

The backgrounds from other \kls decays were reduced to be negligibly small in 
the current experiment, mainly due to the  success of applying tight vetoes with 
a hermetic configuration.  One of the important results is the invariant mass 
distribution of the events with four photons, shown in Fig.~\ref{fig_4gmass}. 
The distribution was well reproduced by the simulations. In particular, the
low-mass region was well described by \kpithrees decays with a small contamination 
by mis-combination events of four photons from \kpitwos decays. The relation between 
\kpithrees and \kpitwos is similar to the relation between \kpitwos and \kznn. 
By taking their branching ratios into account, it was found that the low-mass 
region could to be reproduced after reducing the \kpithrees yield by several orders 
of magnitude.  Reproduction could not be achieved without good simulations of 
small signals from the detector, to which a tight veto was applied.  
This achievement indicates that a similar direction will be promising in the next step.   

The E391a experiment was also able to study other decay modes
including \piz's in the final state, such as 
$K_L^0 \rightarrow \pi^0 \pi^0 \nu \bar{\nu}$ and 
$K_L^0 \rightarrow \pi^0 \pi^0 X$ $(X\rightarrow \gamma \gamma)$.
The \pizs reconstruction method and the hermetic veto system are also valid 
in the analysis of these modes.
Results based on portions of the data have already been published
\cite{ppnn,ppgg}, and further studies are in progress with the entire data set.

The next step, the KOTO experiment \cite{koto} at the new J-PARC accelerator \cite{jparc}, 
is now in preparation.  
Most of the improvements pointed out above are being implemented. 
The new beamline has been constructed; runs to evaluate properties 
of the beamline started in October, 2009.

\begin{acknowledgments}
We are grateful to the crew of the KEK 12-GeV proton synchrotron
for successful beam operation during the experiment. We express our sincere thanks to KEK staff, 
in particular 
the directors: Professors H. Sugawara, Y. Totsuka, A. Suzuki, S. Yamada, M. Kobayashi, 
F. Takasaki, K. Nishikawa, K. Nakai, and K. Nakamura
for their continuous encouragement and support. 
We also express our sincere thanks 
to the many colleagues in universities, institutes, and the kaon-physics
community for their continuous encouragement and support.

This work was partly supported by 
Grant-in-Aids from MEXT and JSPS in Japan, 
grants from NSF and DOE of US, 
grants from NSC in Taiwan,
grants from KRF in Korea,
and the ISTC project from CIS countries.
\\
\end{acknowledgments}

\noindent

\bibliographystyle{plain}

\begin{thebibliography}{99}
\bibitem{e162-csi}
The CsI crystals used in the E162 experiment at KEK-PS were recycled.
\bibitem{ktev-csi}
The CsI crystals were borrowed from the KTeV experiment at Fermilab.
\end{thebibliography}
\noindent
$^a$Present address:  Laboratory of Nuclear Problem, Joint Institute for Nuclear Research, 
Dubna, Moscow Region, 141980 Russia \\
$^b$Present address: KEK, Tsukuba, Ibaraki, 305-0801 Japan. \\
$^c$Present address: CERN, CH-1211 Geneva 23, Switzerland. \\
$^d$Deceased \\
$^e$Present address:  SPring-8, Japan. \\

\end{document}